\documentclass[12pt]{article}
\usepackage{amsmath,amssymb,amsthm,amsxtra,overpic,bbm,bm,epsfig,ulem}
\usepackage{color}
\usepackage{cite}

\usepackage{hyperref}
\usepackage{url}

\def\thefootnote{\fnsymbol{footnote}}
\allowdisplaybreaks[4]

\begin{document}

\vspace{0.2cm}

\begin{center}
{\large\bf Mapping the sources of CP violation in neutrino oscillations \\
from the seesaw mechanism}
\end{center}

\vspace{0.2cm}

\begin{center}
{\bf Zhi-zhong Xing$^{1,2,3}$}
\footnote{E-mail: xingzz@ihep.ac.cn}
\\
{$^{1}$Institute of High Energy Physics,
Chinese Academy of Sciences, Beijing 100049, China \\
$^{2}$School of Physical Sciences,
University of Chinese Academy of Sciences, Beijing 100049, China \\
$^{3}$Center of High Energy Physics, Peking University, Beijing 100871, China}
\end{center}

\vspace{2cm}
\begin{abstract}
We present the first complete calculation of the Jarlskog invariant, a working measure
of the strength of CP violation in the flavor oscillations of three light neutrino
species, with the help of a full Euler-like block parametrization of the flavor
structure in the canonical seesaw mechanism. We find that this invariant depends on
240 linear combinations of the 6 original phase parameters that are responsible for
CP violation in the decays of three heavy Majorana neutrinos in 27 linear combinations
as a whole, and thus provides the first model-independent connection between the
microscopic and macroscopic matter-antimatter asymmetries.
\end{abstract}

\vspace{2cm}


\newpage

\def\thefootnote{\arabic{footnote}}
\setcounter{footnote}{0}

\section{Motivation}

The most natural and most economical extension of the standard model (SM) of
particle physics to generate tiny masses for the three active
neutrinos, whose flavor and mass eigenstates are denoted respectively
as $\nu^{}_\alpha$ (for $\alpha = e, \mu, \tau$) and $\nu^{}_i$ (for
$i = 1, 2, 3$), is to (a) introduce three right-handed neutrino fields
$N^{}_{\alpha \rm R}$ in correspondence to $\nu^{}_{\alpha \rm L}$,
(b) allow the Yukawa interactions between the SM Higgs doublet and the chiral
neutrino fields, and (c) permit the harmless Majorana mass term~\cite{Majorana:1937vz}
formed by the fields $N^{}_{\alpha \rm R}$ and their charge-conjugated states
$(N^{}_{\alpha \rm R})^c$~\cite{Minkowski:1977sc,Yanagida:1979as,GellMann:1980vs,
Glashow:1979nm,Mohapatra:1979ia}. This widely accepted seesaw mechanism
constitutes a canonical theory for the origin of neutrino masses,
and it is fully consistent with the spirit of the SM effective field theory
after the relevant heavy degrees of freedom are integrated out~\cite{Weinberg:1979sa}.
Furthermore, the lepton-number-violating and CP-violating decays of three heavy
Majorana neutrino mass eigenstates $N^{}_j$ (for $j = 4, 5, 6$) give rise to
the well-known leptogenesis mechanism~\cite{Fukugita:1986hr} which can help
explain the observed baryon-antibaryon asymmetry of the
Universe~\cite{ParticleDataGroup:2022pth} in a rather natural way. So the seesaw
mechanism offers a particularly appealing and convincing
killing-two-birds-with-one-stone landscape for the fundamental
parts of both particle physics and cosmology~\cite{Xing:2020ijf}.

The aforementioned seesaw mechanism consists of 18 free parameters: 3 heavy
Majorana neutrino masses $M^{}_j$ (for $j = 4, 5, 6$), 9
active-sterile flavor mixing angles $\theta^{}_{ij}$ (for $i = 1, 2, 3$ and
$j = 4, 5, 6$) and 6 independent CP-violating phases $\alpha^{}_i$ and
$\beta^{}_i$ (for $i = 1, 2, 3$), where $\theta^{}_{ij}$, $\alpha^{}_i$ and
$\beta^{}_i$ are the Euler-like parameters used to describe the seesaw flavor
structure~\cite{Xing:2007zj,Xing:2011ur}. Except $M^{}_i$,
which are expected to be far above the scale of electroweak symmetry breaking,
the other 15 parameters all appear in the leptonic weak charged-current
interactions:
\begin{align}
-{\cal L}^{}_{\rm cc} = & \hspace{0.1cm} \frac{g}{\sqrt{2}} \hspace{0.1cm}
\overline{\big(\begin{matrix} e & \mu & \tau\end{matrix}\big)^{}_{\rm L}}
\hspace{0.1cm} \gamma^\mu \left[ U \left( \begin{matrix} \nu^{}_{1}
\cr \nu^{}_{2} \cr \nu^{}_{3} \cr\end{matrix} \right)^{}_{\hspace{-0.08cm} \rm L}
+ R \left(\begin{matrix} N^{}_4 \cr N^{}_5 \cr N^{}_6
\cr\end{matrix}\right)^{}_{\hspace{-0.08cm} \rm L} \hspace{0.05cm} \right] W^-_\mu
+ {\rm h.c.} \; ,
\label{1}
\end{align}
in which $U = A U^{}_0$ denotes the $3\times 3$ Pontecorvo-Maki-Nakagawa-Sakata
(PMNS) lepton flavor mixing matrix~\cite{Pontecorvo:1957cp,Maki:1962mu,
Pontecorvo:1967fh} with $U^{}_0$ being a unitary matrix, and $R$ measures
the strengths of Yukawa interactions of massive neutrinos and satisfies the
unitarity condition $U U^\dagger + R R^\dagger = A A^\dagger + R R^\dagger = I$.
Note that $A$ and $R$ contain the same active-sterile flavor mixing and CP
violation parameters, and $U$ is also correlated with $R$ via the exact seesaw
relation $U D^{}_\nu U^T = -R D^{}_N R^T$~\cite{Xing:2007zj,Xing:2011ur},
where $D^{}_N \equiv {\rm Diag}\{M^{}_4 , M^{}_5 , M^{}_6\}$ and
$D^{}_\nu \equiv {\rm Diag}\{m^{}_1 , m^{}_2 , m^{}_3\}$ with $m^{}_i$
being the resulting light Majorana neutrino masses (for $i = 1, 2, 3$).
In this case the 9 light degrees of freedom of $D^{}_\nu$ and $U^{}_0$ (i.e.,
the 3 neutrino masses, 3 flavor mixing angles and 3 CP-violating phases)
are {\it derivational} in the sense they can be fully determined
from the 18 {\it original} seesaw flavor parameters~\cite{Xing:2023adc},
\begin{align}
U^{}_0 D^{}_\nu U^T_0 = - \left(A^{-1} R\right) D^{}_N
\left(A^{-1} R\right)^T \; .
\label{2}
\end{align}
In other words, Eq.~(\ref{2}) offers a feasible way to test the seesaw
mechanism at low energies no matter how insignificant the non-unitarity of
$U$ (i.e., the deviation of $U$ from $U^{}_0$) is
\footnote{One may take account of slight quantum corrections between the seesaw
and electroweak scales by means of the renormalization-group equations of the
relevant flavor parameters~\cite{Ohlsson:2013xva,Wang:2023bdw,Zhang:2024weq}.}.

The phase parameters hidden in $A^{-1} R$ are responsible for CP violation in
three heavy Majorana neutrino decays at high energies and must manifest themselves
in the flavor oscillations of three light neutrinos such as $\nu^{}_\mu \to \nu^{}_e$
and $\overline{\nu}^{}_\mu \to \overline{\nu}^{}_e$. It is the unique Jarlskog
invariant~\cite{Jarlskog:1985ht,Wu:1985ea}
\footnote{To avoid any possible confusion, we use $(i, i^\prime, i^{\prime\prime})$
to denote the {\it light} neutrino indices $(1, 2, 3)$, and
$(j, k, l)$ to denote the {\it heavy} neutrino indices $(4, 5, 6)$
throughout this paper.}
\begin{align}
{\cal J}^{}_\nu \equiv {\rm Im}\left[\left(U^{}_0\right)^{}_{\alpha i}
\left(U^{}_0\right)^{}_{\beta i^\prime} \left(U^{}_0\right)^{*}_{\alpha i^\prime}
\left(U^{}_0\right)^{*}_{\beta i} \right] \; ,
\label{3}
\end{align}
in which the Greek subscripts $(\alpha, \beta)$ run cyclically over $(e, \mu, \tau)$
and the Latin subscripts $(i, i^\prime)$ run cyclically over $(1, 2, 3)$, that
universally measures the strength of CP-violating effects in neutrino oscillations
in the $U \to U^{}_0$ limit. Given the fact that $U \simeq U^{}_0$ is an excellent
approximation (i.e., the non-unitarity of $U$ has been well constrained from a
global analysis of current data on flavor and electroweak precision measurements~\cite{Antusch:2014woa,Blennow:2016jkn,Wang:2021rsi,Blennow:2023mqx}),
${\cal J}^{}_\nu$ is expected to be a working description of leptonic
CP violation at low energies to a good degree of accuracy. The T2K
long-baseline neutrino oscillation experiment has so far ruled out
${\cal J}^{}_\nu = 0$ at the $2\sigma$ level~\cite{T2K:2019bcf,T2K:2023smv}.

So a model-independent calculation of ${\cal J}^{}_\nu$ in terms of the 6
original CP-violating phases in the seesaw mechanism is fundamentally important
to establish a testable correlation between the microscopic and macroscopic
matter-antimatter asymmetries. But such a calculation has been lacking,
as a general and direct connection between the effects of CP
violation at high and low energies was concluded to be impossible in the
seesaw framework due to its unknown flavor structure~\cite{Buchmuller:1996pa}.

We are going to solve this long-standing problem by presenting the first complete
calculation of the Jarlskog invariant ${\cal J}^{}_\nu$ with the help of a
full Euler-like block parametrization of the seesaw flavor texture proposed
in Refs.~\cite{Xing:2007zj,Xing:2011ur}. We demonstrate that ${\cal J}^{}_\nu$
depends directly on 240 linear combinations of the 6 original phase parameters in
$R$, which are responsible for CP violation in the lepton-number-violating decays
of three heavy Majorana neutrinos in 27 linear combinations as a whole and thus for
the validity of leptogenesis. This model-independent result will help open an
important bottom-up window to probe the cosmic matter-antimatter asymmetry from
CP violation in neutrino oscillations at low energies via the canonical seesaw bridge.

\section{Calculations of ${\cal J}^{}_\nu$}

Taking account of the Euler-like block parametrization of the $6\times 6$ unitary
flavor mixing matrix in the canonical seesaw mechanism~\cite{Xing:2011ur}, we have
the exact expressions of $A$ and $R$ as follows:
\begin{align}
A = & \left( \begin{matrix} c^{}_{14} c^{}_{15} c^{}_{16} & 0 & 0
\cr \vspace{-0.45cm} \cr
\begin{array}{l} -c^{}_{14} c^{}_{15} \hat{s}^{}_{16} \hat{s}^*_{26} -
c^{}_{14} \hat{s}^{}_{15} \hat{s}^*_{25} c^{}_{26} \\
-\hat{s}^{}_{14} \hat{s}^*_{24} c^{}_{25} c^{}_{26} \end{array} &
c^{}_{24} c^{}_{25} c^{}_{26} & 0 \cr \vspace{-0.45cm} \cr
\begin{array}{l} -c^{}_{14} c^{}_{15} \hat{s}^{}_{16} c^{}_{26} \hat{s}^*_{36}
+ c^{}_{14} \hat{s}^{}_{15} \hat{s}^*_{25} \hat{s}^{}_{26} \hat{s}^*_{36} \\
- c^{}_{14} \hat{s}^{}_{15} c^{}_{25} \hat{s}^*_{35} c^{}_{36} +
\hat{s}^{}_{14} \hat{s}^*_{24} c^{}_{25} \hat{s}^{}_{26}
\hat{s}^*_{36} \\
+ \hat{s}^{}_{14} \hat{s}^*_{24} \hat{s}^{}_{25} \hat{s}^*_{35}
c^{}_{36} - \hat{s}^{}_{14} c^{}_{24} \hat{s}^*_{34} c^{}_{35}
c^{}_{36} \end{array} &
\begin{array}{l}
\hspace{0.04cm} -c^{}_{24} c^{}_{25} \hat{s}^{}_{26} \hat{s}^*_{36} -
c^{}_{24} \hat{s}^{}_{25} \hat{s}^*_{35} c^{}_{36} \hspace{0.04cm} \\
\hspace{0.04cm} -\hat{s}^{}_{24} \hat{s}^*_{34} c^{}_{35} c^{}_{36} \end{array} &
c^{}_{34} c^{}_{35} c^{}_{36} \cr \end{matrix} \right) \; ,
\label{4}
\end{align}
and
\begin{align}
R = & \left( \begin{matrix} \hat{s}^*_{14} c^{}_{15} c^{}_{16} &
\hat{s}^*_{15} c^{}_{16} & \hat{s}^*_{16} \cr \vspace{-0.45cm} \cr
\begin{array}{l} -\hat{s}^*_{14} c^{}_{15} \hat{s}^{}_{16} \hat{s}^*_{26} -
\hat{s}^*_{14} \hat{s}^{}_{15} \hat{s}^*_{25} c^{}_{26} \\
+ c^{}_{14} \hat{s}^*_{24} c^{}_{25} c^{}_{26} \end{array} & -
\hat{s}^*_{15} \hat{s}^{}_{16} \hat{s}^*_{26} + c^{}_{15}
\hat{s}^*_{25} c^{}_{26} & c^{}_{16} \hat{s}^*_{26} \cr \vspace{-0.45cm} \cr
\begin{array}{l} -\hat{s}^*_{14} c^{}_{15} \hat{s}^{}_{16} c^{}_{26}
\hat{s}^*_{36} + \hat{s}^*_{14} \hat{s}^{}_{15} \hat{s}^*_{25}
\hat{s}^{}_{26} \hat{s}^*_{36} \\ - \hat{s}^*_{14} \hat{s}^{}_{15}
c^{}_{25} \hat{s}^*_{35} c^{}_{36} - c^{}_{14} \hat{s}^*_{24}
c^{}_{25} \hat{s}^{}_{26}
\hat{s}^*_{36} \\
- c^{}_{14} \hat{s}^*_{24} \hat{s}^{}_{25} \hat{s}^*_{35}
c^{}_{36} + c^{}_{14} c^{}_{24} \hat{s}^*_{34} c^{}_{35} c^{}_{36}
\end{array} &
\begin{array}{l} -\hat{s}^*_{15} \hat{s}^{}_{16} c^{}_{26} \hat{s}^*_{36}
- c^{}_{15} \hat{s}^*_{25} \hat{s}^{}_{26} \hat{s}^*_{36} \\
+c^{}_{15} c^{}_{25} \hat{s}^*_{35} c^{}_{36} \end{array} &
c^{}_{16} c^{}_{26} \hat{s}^*_{36} \cr \end{matrix} \right) \; , \hspace{0.5cm}
\label{5}
\end{align}
where $c^{}_{ij} \equiv \cos\theta^{}_{ij}$, $s^{}_{ij} \equiv \sin\theta^{}_{ij}$
and $\hat{s}^{}_{ij} \equiv s^{}_{ij} e^{{\rm i}\delta^{}_{ij}}$ with
$\theta^{}_{ij}$ lying in the first quadrant. As a consequence,
\begin{align}
A^{-1} R = \left( \begin{matrix} \hat{t}^{*}_{14} & c^{-1}_{14} \hat{t}^{*}_{15} &
c^{-1}_{14} c^{-1}_{15} \hat{t}^{*}_{16}
\cr \vspace{-0.45cm} \cr
c^{-1}_{14} \hat{t}^{*}_{24} &
\hat{t}^{}_{14} \hat{t}^{*}_{15} \hat{t}^{*}_{24} + c^{-1}_{15} c^{-1}_{24}
\hat{t}^{*}_{25} &
\begin{array}{l}
+ \hat{t}^{}_{14} c^{-1}_{15} \hat{t}^{*}_{16} \hat{t}^{*}_{24}
+ \hat{t}^{}_{15} \hat{t}^{*}_{16} c^{-1}_{24} \hat{t}^{*}_{25} \\
+ c^{-1}_{16} c^{-1}_{24} c^{-1}_{25} \hat{t}^{*}_{26} \end{array}
\cr \vspace{-0.45cm} \cr
c^{-1}_{14} c^{-1}_{24} \hat{t}^{*}_{34} &
\begin{array}{l}
+ \hat{t}^{}_{14} \hat{t}^{*}_{15} c^{-1}_{24} \hat{t}^{*}_{34}
+ c^{-1}_{15} \hat{t}^{}_{24} \hat{t}^{*}_{25} \hat{t}^{*}_{34} \\
+ c^{-1}_{15} c^{-1}_{25} c^{-1}_{34} \hat{t}^{*}_{35}
\end{array} &
\begin{array}{l}
+ \hat{t}^{}_{14} c^{-1}_{15} \hat{t}^{*}_{16} c^{-1}_{24} \hat{t}^{*}_{34}
+ \hat{t}^{}_{15} \hat{t}^{*}_{16} \hat{t}^{}_{24} \hat{t}^{*}_{25}
\hat{t}^{*}_{34} \\
+ \hat{t}^{}_{15} \hat{t}^{*}_{16} c^{-1}_{25} c^{-1}_{34} \hat{t}^{*}_{35}
+ c^{-1}_{16} \hat{t}^{}_{24} c^{-1}_{25} \hat{t}^{*}_{26} \hat{t}^{*}_{34} \\
+ c^{-1}_{16} \hat{t}^{}_{25} \hat{t}^{*}_{26} c^{-1}_{34} \hat{t}^{*}_{35}
+ c^{-1}_{16} c^{-1}_{26} c^{-1}_{34} c^{-1}_{35} \hat{t}^{*}_{36}
\end{array}
\cr \end{matrix} \right) \; ,
\label{6}
\end{align}
where $\hat{t}^{}_{ij} \equiv e^{{\rm i} \delta^{}_{ij}} \tan\theta^{}_{ij}$
is defined. Some constraints on the 9 active-sterile flavor mixing angles
$\theta^{}_{ij}$ (for $i = 1, 2, 3$ and $j = 4, 5, 6$) have been obtained from the
currently available bounds on non-unitarity of the PMNS lepton flavor mixing
matrix $U$~\cite{Antusch:2014woa,Blennow:2016jkn,Wang:2021rsi,Blennow:2023mqx}
via the relations $A A^\dagger = U U^\dagger$ and $R R^\dagger = I - A A^\dagger$,
from which one finds that the magnitudes of $\theta^{}_{ij}$ must be smaller
than ${\cal O}(10^{-1.5})$. So a Maclaurin expansion of Eq.~(\ref{6}) leads
us to
\begin{align}
A^{-1} R =
\left(\begin{matrix} \hat{s}^*_{14} & \hat{s}^*_{15} & \hat{s}^*_{16} \cr
\hat{s}^*_{24} & \hat{s}^*_{25} & \hat{s}^*_{26} \cr
\hat{s}^*_{34} & \hat{s}^*_{35} & \hat{s}^*_{36} \cr \end{matrix}\right)
+ {\cal O}(s^3_{ij}) + \cdots \; .
\label{7}
\end{align}
Keeping only the leading term of $A^{-1} R$ turns out to be a very good
approximation, independent of any possible hierarchy or degeneracy of the
original seesaw parameters. The smallness of active-sterile flavor
mixing implies that the general Jarlskog invariants of CP violation
defined in terms of $U$ is actually dominated by the unique ${\cal J}^{}_\nu$
defined in Eq.~(\ref{3}) with the help of $U^{}_0$, namely
\begin{align}
{\cal J}^{i i^\prime}_{\alpha\beta} \equiv {\rm Im}\left(U^{}_{\alpha i}
U^{}_{\beta i^\prime} U^{*}_{\alpha i^\prime} U^{*}_{\beta i} \right) =
{\cal J}^{}_\nu + \Delta {\cal J}^{i i^\prime}_{\alpha\beta} \; ,
\label{8}
\end{align}
where $(\alpha, \beta)$ run cyclically over $(e, \mu, \tau)$, $(i, i^\prime)$
run cyclically over $(1, 2, 3)$, and $\Delta {\cal J}^{i i^\prime}_{\alpha\beta}$
stand for the corresponding small corrections to ${\cal J}^{}_\nu$.
Taking account of the recent numerical bounds on possible deviations of
$A A^\dagger$ from $I$~\cite{Blennow:2023mqx} and the fact that
$\Delta {\cal J}^{i i^\prime}_{\alpha\beta}$ are at least suppressed by the factors
of ${\cal O} (s^2_{ij})$ (for $i = 1, 2, 3$ and $j = 4, 5, 6$)~\cite{Xing:2011ur},
we find that
$|\Delta {\cal J}^{i i^\prime}_{\alpha\beta}/{\cal J}^{}_\nu| \lesssim 10^{-2}$
is a safe estimate especially for $|{\cal J}^{}_\nu| \sim 10^{-2}$ as favored
by the present T2K data~\cite{T2K:2019bcf,T2K:2023smv}. In other words,
${\cal J}^{i i^\prime}_{\alpha\beta} \simeq {\cal J}^{}_\nu$ is also expected
to be a very good approximation. So it will make sense to substitute the approximate
expression of $A^{-1} R$ in Eq.~(\ref{7}) into the seesaw formula in Eq.~(\ref{2})
to calculate all the leading terms of ${\cal J}^{}_\nu$.

A proper redefinition of
the phases of three charged lepton fields in Eq.~(\ref{1}) allows one to rotate
away 3 of the 9 phase parameters (or their independent combinations) of $A$ and
$R$. We are therefore left with 6 nontrivial CP-violating phases which can
be chosen from
\begin{align}
\alpha^{}_i \equiv \delta^{}_{i 4} - \delta^{}_{i 5} \; , \quad
\beta^{}_i \equiv \delta^{}_{i 5} - \delta^{}_{i 6} \; , \quad
\gamma^{}_i \equiv \delta^{}_{i 6} - \delta^{}_{i 4} \; ,
\label{9}
\end{align}
as they satisfy $\alpha^{}_i + \beta^{}_i + \gamma^{}_i = 0$ (for $i = 1, 2, 3$).
Substituting Eq.~(\ref{7}) into Eq.~(\ref{2}) will make it possible for us
to work out all the flavor parameters of the three light Majorana neutrinos.
But in this work we concentrate on leptonic CP violation --- a burning issue
of particle physics which will definitely be established in the
next-generation long-baseline neutrino oscillation
experiments~\cite{Hyper-KamiokandeProto-:2015xww,DUNE:2015lol}.

We find that the most convenient way to calculate the Jarlskog invariant
${\cal J}^{}_\nu$ defined in Eq.~(\ref{3}) is
\begin{align}
{\cal J}^{}_\nu = \frac{{\rm Im}\left[\left(M^{}_\nu M^\dagger_\nu\right)^{}_{e\mu}
\left(M^{}_\nu M^\dagger_\nu\right)^{}_{\mu\tau}
\left(M^{}_\nu M^\dagger_\nu\right)^{}_{\tau e}\right]}
{\Delta^{}_{21} \Delta^{}_{31} \Delta^{}_{32}} \; ,
\label{10}
\end{align}
where $M^{}_\nu \equiv U^{}_0 D^{}_\nu U^T_0$ and $\Delta^{}_{i i^\prime} \equiv
m^2_i - m^2_{i^\prime}$ (for $i, i^\prime = 1, 2, 3$). One may derive the
expressions of $m^2_i$ by use of the parameters of $D^{}_N$ and $A^{-1} R$ from
Eq.~(\ref{2}), and then arrive at those of $\Delta^{}_{i i^\prime}$.
But a global analysis of the neutrino oscillation data has given the
best-fit values $\Delta^{}_{21} \simeq 7.41 \times 10^{-5} ~{\rm eV}^2$ and
$\Delta^{}_{31} \simeq 2.51 \times 10^{-3} ~{\rm eV}^2$ (normal mass
ordering) or $\Delta^{}_{31} \simeq -2.41 \times 10^{-3} ~{\rm eV}^2$
(inverted mass ordering)~\cite{Gonzalez-Garcia:2021dve,Capozzi:2021fjo}.
So the product $\Delta^{}_{21} \Delta^{}_{31} \Delta^{}_{32}$ on the
right-hand side of Eq.~(\ref{10}) is already known to us, and it is
insensitive to the sign ambiguities of $\Delta^{}_{31}$ and $\Delta^{}_{32}$.
In this case what we need to do is to calculate
$(M^{}_\nu M^\dagger_\nu)^{}_{e\mu}$, $(M^{}_\nu M^\dagger_\nu)^{}_{\mu\tau}$
and $(M^{}_\nu M^\dagger_\nu)^{}_{\tau e}$ in terms of the original seesaw
parameters $M^{}_j$, $\theta^{}_{ij}$ and $\delta^{}_{ij}$ (for
$i = 1, 2, 3$ and $j = 4, 5, 6$) with the help of Eqs.~(\ref{2}) and
(\ref{7}):
\footnote{If the exact expression of $A^{-1} R$ in Eq.~(\ref{6}) is used to
calculate ${\cal J}^{}_\nu$ with the help of Eqs.~(\ref{2}) and (\ref{10}),
much more independent phase combinations are expected to appear in the
sub-leading terms of the analytical expression of ${\cal J}^{}_\nu$.
But such terms are suppressed by the higher-order factors of the small
active-sterile flavor mixing angles, as can be easily seen from Eq.~(\ref{7}),
and hence their effects are negligibly small.}
\begin{align}
\left(M^{}_\nu M^\dagger_\nu\right)^{}_{e\mu} = & \hspace{0.1cm}
\sum^6_{j=4} \sum^6_{k=4} M^{}_j M^{}_k I^{}_{jk} \hat{s}^*_{1j} \hat{s}^{}_{2k} \; ,
\nonumber \\
\left(M^{}_\nu M^\dagger_\nu\right)^{}_{\mu\tau} = & \hspace{0.1cm}
\sum^6_{j=4} \sum^6_{k=4} M^{}_j M^{}_k I^{}_{jk} \hat{s}^*_{2j} \hat{s}^{}_{3k} \; ,
\nonumber \\
\left(M^{}_\nu M^\dagger_\nu\right)^{}_{\tau e} = & \hspace{0.1cm}
\sum^6_{j=4} \sum^6_{k=4} M^{}_j M^{}_k I^{}_{jk} \hat{s}^{*}_{3j} \hat{s}^{}_{1k} \; ,
\label{11}
\end{align}
where
\begin{align}
I^{}_{jk} \equiv \sum^3_{i=1} \hat{s}^*_{ij} \hat{s}^{}_{ik} = I^*_{kj} \; ,
\; \left(j, k = 4, 5, 6\right) \; .
\label{12}
\end{align}
It is obvious that $I^{}_{44}$, $I^{}_{55}$ and $I^{}_{66}$ are real and
positive, while $I^{}_{45}$, $I^{}_{56}$ and $I^{}_{64}$ depend respectively
on the phase parameters $\alpha^{}_i$, $\beta^{}_i$ and $\gamma^{}_i$
(for $i = 1, 2, 3$). Note that $I^{}_{56}$ or $I^{}_{64}$ can easily be
obtained from $I^{}_{45}$ by making the subscript replacements $(4, 5) \to (5, 6)$
or $(4, 5) \to (6, 4)$, together with $\alpha^{}_i \to \beta^{}_i$ or
$\alpha^{}_i \to \gamma^{}_i$. Substituting Eq.~(\ref{11}) into Eq.~(\ref{10})
will then allow us to arrive at the explicit dependence of ${\cal J}^{}_\nu$
on the 6 original CP-violating phases in the seesaw framework.

A very lengthy calculation shows that the terms of
${\rm Im}\left[(M^{}_\nu M^\dagger_\nu)^{}_{e\mu}
(M^{}_\nu M^\dagger_\nu)^{}_{\mu\tau}
(M^{}_\nu M^\dagger_\nu)^{}_{\tau e}\right]$ that are proportional to $M^6_j$
(for $j = 4, 5, 6$) and $M^5_j M^{}_k$ (for $k \neq j$) do not contribute to
${\cal J}^{}_\nu$ at all. But the terms of
${\rm Im}\left[(M^{}_\nu M^\dagger_\nu)^{}_{e\mu}
(M^{}_\nu M^\dagger_\nu)^{}_{\mu\tau}
(M^{}_\nu M^\dagger_\nu)^{}_{\tau e}\right]$ proportional to
$M^3_j M^3_k$ and $M^4_j M^2_k$ (for $k \neq j$) contribute to
${\cal J}^{}_\nu$, so do the terms proportional to $M^4_j M^{}_k M^{}_l$,
$M^3_j M^2_k M^{}_l$ and $M^2_j M^2_k M^2_l$ (for $l \neq k \neq j$).
So we proceed to present the analytical results of those nontrivial
terms.

\subsection{Terms $T^{}_{\bf 33}$ and $T^{}_{\bf 42}$}

For the sake of simplicity, we denote a sum of the terms of
${\rm Im}\left[(M^{}_\nu M^\dagger_\nu)^{}_{e\mu} (M^{}_\nu
M^\dagger_\nu)^{}_{\mu\tau} (M^{}_\nu M^\dagger_\nu)^{}_{\tau e}\right]$
proportional to $M^3_j M^3_k$ (for $k \neq j$) as
$T^{}_{\bf 33}$, and a sum of those terms proportional to $M^4_j M^2_k$ (for
$k \neq j$) as $T^{}_{\bf 42}$. Our results of $T^{}_{\bf 33}$ and
$T^{}_{\bf 42}$ are summarized as
\begin{align}
T^{}_{\bf 33} = & \hspace{0.1cm}
M^3_4 M^3_5 \left(I^{}_{44} I^{}_{55} - \left|I^{}_{45}\right|^2\right)
\Bigg[\sum^3_{i=1} s^{}_{i4} s^{}_{i5} s^{}_{i^\prime 4}
s^{}_{i^\prime 5} \left(s^2_{i4} s^2_{i^{\prime\prime} 5}
+ s^2_{i^{\prime\prime} 4} s^2_{i^\prime 5}
- s^2_{i^\prime 4} s^2_{i^{\prime\prime} 5}
- s^2_{i^{\prime\prime} 4} s^2_{i5}\right)
\sin\left(\alpha^{}_i + \alpha^{}_{i^\prime}\right)
\nonumber \\
& - \hspace{0.08cm} \sum^3_{i=1} s^2_{i4} s^2_{i5}
\left(s^2_{i^\prime 4} s^2_{i^{\prime\prime} 5} -
s^2_{i^{\prime\prime} 4} s^2_{i^\prime 5}\right) \sin 2\alpha^{}_i
\Bigg]
\nonumber \\
& + {\rm term}\big\{4 \to 5, \; 5 \to 6; \;
\alpha^{}_i \to \beta^{}_i\big\}
+ {\rm term}\big\{4 \to 6, \; 5 \to 4; \;
\alpha^{}_i \to \gamma^{}_i\big\} \; ,
\label{13}
\\
T^{}_{\bf 42} = & \hspace{0.1cm}
M^2_4 M^2_5 \left(I^{}_{44} I^{}_{55} - \left|I^{}_{45}\right|^2\right)
\Bigg[\sum^3_{i=1} s^{}_{i4} s^{}_{i5} s^{}_{i^\prime 4} s^{}_{i^\prime 5}
\left(M^2_4 I^{}_{44} s^2_{i^{\prime\prime} 4}
- M^2_5 I^{}_{55} s^2_{i^{\prime\prime} 5}\right)
\sin\left(\alpha^{}_i - \alpha^{}_{i^\prime}\right)\Bigg]
\nonumber \\
& + {\rm term}\big\{4 \to 5, \; 5 \to 6; \;
\alpha^{}_i \to \beta^{}_i\big\}
+ {\rm term}\big\{4 \to 6, \; 5 \to 4; \;
\alpha^{}_i \to \gamma^{}_i\big\} \; ,
\label{14}
\end{align}
where the subscripts $i$, $i^\prime$ and $i^{\prime\prime}$ run
cyclically over $1$, $2$ and $3$, the
``term~$\left\{4 \to 5, \; 5 \to 6; \; \alpha^{}_i \to \beta^{}_i\right\}$"
means that another term of $T^{}_{\bf 33}$ or $T^{}_{\bf 42}$ can be easily
written out from the above-obtained expression by making the {\it heavy} subscript
replacements $4 \to 5$ and $5 \to 6$ together with $\alpha^{}_i \to \beta^{}_i$
(for $i = 1, 2, 3$), and so is the case for the abbreviation
``term~$\left\{4 \to 6, \; 5 \to 4; \; \alpha^{}_i \to \gamma^{}_i\right\}$".
It is obvious that $T^{}_{\bf 33}$ and $T^{}_{\bf 42}$ characterize the
two-species interference effects among the three species of heavy Majorana
neutrinos, and hence their expressions are relatively simple.

\subsection{Term $T^{}_{\bf 411}$}

The terms of ${\rm Im}\left[(M^{}_\nu M^\dagger_\nu)^{}_{e\mu}
(M^{}_\nu M^\dagger_\nu)^{}_{\mu\tau} (M^{}_\nu M^\dagger_\nu)^{}_{\tau e}\right]$
proportional to $M^4_j M^{}_k M^{}_l$ (for $l \neq k \neq j$) are summarily denoted as
$T^{}_{\bf 411}$. The explicit analytical result of $T^{}_{\bf 411}$ is found to be
\begin{align}
T^{}_{\bf 411} = & \hspace{0.13cm}
M^4_4 M^{}_5 M^{}_6 I^{}_{44} s^{}_{14} s^{}_{24} s^{}_{34} \Bigg[
- s^{}_{14} s^{}_{24} s^{}_{34} \sum^3_{i=1} s^2_{i5}
\Big[s^2_{i^\prime 6} \sin 2\left(\alpha^{}_i + \gamma^{}_{i^\prime}\right)
- s^2_{i^{\prime\prime} 6} \sin 2\left(\alpha^{}_i
+ \gamma^{}_{i^{\prime\prime}}\right)\Big]
\nonumber \\
&
+ \sum^3_{i=1} s^{}_{i4} \left(s^2_{i^\prime 4} - s^2_{i^{\prime\prime} 4}
\right) \Big[ s^2_{i5} s^{}_{i^\prime 6} s^{}_{i^{\prime\prime} 6}
\sin\left(2 \alpha^{}_i + \gamma^{}_{i^\prime} + \gamma^{}_{i^{\prime\prime}}
\right)
- s^2_{i6} s^{}_{i^\prime 5} s^{}_{i^{\prime\prime} 5}
\sin\left(\alpha^{}_{i^\prime} + \alpha^{}_{i^{\prime\prime}}
+ 2\gamma^{}_i\right) \Big]
\nonumber \\
&
+ \sum^3_{i=1} s^{}_{i4} s^{}_{i^\prime 5} s^{}_{i^\prime 6} \left(s^2_{i4}
+ s^2_{i^{\prime\prime} 4}\right)
\Big[ s^{}_{i^\prime 5} s^{}_{i^{\prime\prime} 6}
\sin\left(\alpha^{}_{i^\prime} - \beta^{}_{i^\prime}
+ \gamma^{}_{i^{\prime\prime}}\right) - s^{}_{i^\prime 6} s^{}_{i^{\prime\prime} 5}
\sin\left(\alpha^{}_{i^{\prime\prime}} - \beta^{}_{i^\prime}
+ \gamma^{}_{i^{\prime}} \right)\Big]
\nonumber \\
&
+ \sum^3_{i=1} s^{}_{i4} s^{}_{i^{\prime\prime} 5} s^{}_{i^{\prime\prime} 6}
\left(s^2_{i4} + s^2_{i^{\prime} 4}\right)
\Big[ s^{}_{i^\prime 5} s^{}_{i^{\prime\prime} 6}
\sin\left(\alpha^{}_{i^\prime} - \beta^{}_{i^{\prime\prime}}
+ \gamma^{}_{i^{\prime\prime}}\right) - s^{}_{i^\prime 6} s^{}_{i^{\prime\prime} 5}
\sin\left(\alpha^{}_{i^{\prime\prime}} - \beta^{}_{i^{\prime\prime}}
+ \gamma^{}_{i^{\prime}} \right)\Big]
\nonumber \\
&
+ \sum^3_{i=1} 2 s^{}_{i4} s^{}_{i5} s^{}_{i6} \left(s^2_{i^\prime 4}
+ s^2_{i^{\prime\prime} 4}\right)
\Big[ s^{}_{i^\prime 5} s^{}_{i^{\prime\prime} 6}
\sin\left(\alpha^{}_{i^\prime} - \beta^{}_i + \gamma^{}_{i^{\prime\prime}}\right)
- s^{}_{i^\prime 6} s^{}_{i^{\prime\prime} 5}
\sin\left(\alpha^{}_{i^{\prime\prime}} - \beta^{}_i + \gamma^{}_{i^{\prime}}
\right)\Big] \Bigg]
\nonumber \\
&
+ {\rm term}\big\{\left(4, 5, 6\right) \to \left(5, 4, 6\right) ; \;
\left(\alpha^{}_i, \;\beta^{}_i, \;\gamma^{}_i\right) \to
-\left(\alpha^{}_i, \;\gamma^{}_i, \;\beta^{}_i\right)\big\}
\nonumber \\
&
+ {\rm term}\big\{\left(4, 5, 6\right) \to \left(6, 5, 4\right) ; \;
\left(\alpha^{}_i, \;\beta^{}_i, \;\gamma^{}_i\right) \to
-\left(\beta^{}_i, \;\alpha^{}_i, \;\gamma^{}_i\right)\big\} \; ,
\label{15}
\end{align}
where the three {\it light} subscripts $\left(i, i^\prime, i^{\prime\prime}\right)$
run cyclically over $\left(1, 2, 3\right)$, and the {\it heavy} subscript replacements
for those abbreviation terms have the same implications as in Eqs.~(\ref{13}) and (\ref{14}).

\subsection{Term $T^{}_{\bf 321}$}

Let us denote a sum of the terms of ${\rm Im}\left[(M^{}_\nu M^\dagger_\nu)^{}_{e\mu}
(M^{}_\nu M^\dagger_\nu)^{}_{\mu\tau} (M^{}_\nu M^\dagger_\nu)^{}_{\tau e}\right]$
proportional to $M^3_4 M^2_5 M^{}_6$, $M^3_4 M^2_6 M^{}_5$, $M^3_5 M^2_4 M^{}_6$,
$M^3_5 M^2_6 M^{}_4$, $M^3_6 M^2_4 M^{}_5$ and $M^3_6 M^2_5 M^{}_4$ as
$T^{}_{\bf 321}$. The explicit analytical result of $T^{}_{\bf 321}$
is obtained as follows:
\begin{align}
T^{}_{\bf 321} = & \hspace{0.13cm}
M^3_4 M^2_5 M^{}_6 \Bigg\{
\left(I^{}_{44} I^{}_{55} - |I^{}_{45}|^2\right) \Bigg[
\sum^3_{i=1} s^2_{i4} s^{}_{i^\prime 5} s^{}_{i^{\prime\prime} 5}
\Big[s^{}_{i4} s^{}_{i6} s^{}_{i^\prime 4} s^{}_{i^{\prime\prime} 6}
\sin\left(\alpha^{}_{i^\prime} + \beta^{}_{i^{\prime\prime}} - \gamma^{}_i\right)
- s^{}_{i4} s^{}_{i6} s^{}_{i^\prime 6} s^{}_{i^{\prime\prime} 4}
\nonumber \\
&
\times \sin\left(\alpha^{}_{i^{\prime\prime}}
+ \beta^{}_{i^{\prime}} - \gamma^{}_i\right)
+ s^2_{i^\prime 4} s^{}_{i^\prime 6} s^{}_{i^{\prime\prime} 6}
\sin\left(\alpha^{}_{i^\prime} + \beta^{}_{i^{\prime\prime}} -
\gamma^{}_{i^\prime}\right)
- s^{}_{i^\prime 4} s^2_{i^\prime 6} s^{}_{i^{\prime\prime} 4}
\sin\left(\alpha^{}_{i^{\prime\prime}} + \beta^{}_{i^{\prime}}
- \gamma^{}_{i^\prime}\right)
\nonumber \\
&
+ s^{}_{i^\prime 4} s^{}_{i^{\prime\prime} 4} s^2_{i^{\prime\prime} 6}
\sin\left(\alpha^{}_{i^\prime} + \beta^{}_{i^{\prime\prime}}
- \gamma^{}_{i^{\prime\prime}}\right)
- s^{}_{i^\prime 6} s^2_{i^{\prime\prime} 4} s^{}_{i^{\prime\prime} 6}
\sin\left(\alpha^{}_{i^{\prime\prime}} + \beta^{}_{i^{\prime}}
- \gamma^{}_{i^{\prime\prime}}\right)
\Big] \Bigg]
\nonumber \\
&
+ I^{}_{44} s^{}_{14} s^{}_{24} s^{}_{34} \Bigg[
2 \sum^3_{i=1} s^{}_{i 6} s^{}_{i^\prime 5} s^{}_{i^{\prime\prime} 5}
\Big[ s^{}_{i4} s^{}_{i5} s^{}_{i^\prime 5} s^{}_{i^\prime 6}
\cos\left(\alpha^{}_i + \beta^{}_{i^\prime} - \gamma^{}_i\right)
+ s^{}_{i4} s^{}_{i5} s^{}_{i^{\prime\prime} 5} s^{}_{i^{\prime\prime} 6}
\nonumber \\
&
\times \cos\left(\alpha^{}_i + \beta^{}_{i^{\prime\prime}} - \gamma^{}_i\right)
+ s^{}_{i^\prime 4} s^{}_{i^\prime 5} s^{}_{i 5} s^{}_{i 6}
\cos\left(\alpha^{}_{i^\prime} + \beta^{}_i - \gamma^{}_i\right)
+ s^{}_{i^\prime 4} s^{}_{i^\prime 5} s^{}_{i^{\prime\prime} 5} s^{}_{i^{\prime\prime} 6}
\cos\left(\alpha^{}_{i^{\prime}} + \beta^{}_{i^{\prime\prime}} - \gamma^{}_i\right)
\nonumber\\
&
+ s^{}_{i^{\prime\prime} 4} s^{}_{i^{\prime\prime} 5} s^{}_{i 5} s^{}_{i 6}
\cos\left(\alpha^{}_{i^{\prime\prime}} + \beta^{}_i - \gamma^{}_i\right)
+ s^{}_{i^{\prime\prime} 4} s^{}_{i^{\prime\prime} 5} s^{}_{i^{\prime} 5}
s^{}_{i^{\prime} 6} \cos\left(\alpha^{}_{i^{\prime\prime}} + \beta^{}_{i^{\prime}}
- \gamma^{}_i\right)
\nonumber\\
&
- s^{}_{i4} s^{}_{i6} \left(s^2_{i^\prime 5} + s^2_{i^{\prime\prime} 5}
\right) \cos 2\gamma^{}_i - s^{}_{i^\prime 4} s^{}_{i^\prime 6}
\left(s^2_{i 5} + s^2_{i^{\prime\prime} 5}\right) \cos\left(\gamma^{}_i
+ \gamma^{}_{i^\prime}\right)
\nonumber \\
&
- s^{}_{i^{\prime\prime} 4} s^{}_{i^{\prime\prime} 6}
\left(s^2_{i 5} + s^2_{i^{\prime} 5}\right) \cos\left(\gamma^{}_i +
\gamma^{}_{i^{\prime\prime}}\right)\Big]
\sin\left(\alpha^{}_{i^{\prime\prime}} - \alpha^{}_{i^\prime}\right)\Bigg]
\nonumber \\
&
+ I^{}_{44} \Bigg[
\sum^3_{i=1} s^{}_{i4} s^{}_{i6} \left(s^2_{i^\prime 5} s^2_{i^{\prime\prime} 4}
- s^2_{i^\prime 4} s^2_{i^{\prime\prime} 5}\right)
\Big[s^{}_{i4} s^{}_{i5} s^{}_{i^\prime 5} s^{}_{i^\prime 6}
\sin\left(\alpha^{}_i + \beta^{}_{i^\prime} - \gamma^{}_i\right)
+ s^{}_{i4} s^{}_{i5} s^{}_{i^{\prime\prime} 5} s^{}_{i^{\prime\prime} 6}
\nonumber \\
&
\times \sin\left(\alpha^{}_i + \beta^{}_{i^{\prime\prime}} - \gamma^{}_i\right)
+ s^{}_{i^\prime 4} s^{}_{i^\prime 5} s^{}_{i^{\prime\prime} 5} s^{}_{i^{\prime\prime} 6}
\sin\left(\alpha^{}_{i^\prime} + \beta^{}_{i^{\prime\prime}} - \gamma^{}_i\right)
+ s^{}_{i^\prime 4} s^{}_{i^\prime 5} s^{}_{i 5} s^{}_{i 6}
\sin\left(\alpha^{}_{i^\prime} + \beta^{}_i - \gamma^{}_i\right)
\nonumber \\
&
+ s^{}_{i^{\prime\prime} 4} s^{}_{i^{\prime\prime} 5} s^{}_{i 5} s^{}_{i 6}
\sin\left(\alpha^{}_{i^{\prime\prime}} + \beta^{}_i - \gamma^{}_i\right)
+ s^{}_{i^{\prime\prime} 4} s^{}_{i^{\prime\prime} 5} s^{}_{i^\prime 5}
s^{}_{i^\prime 6} \sin\left(\alpha^{}_{i^{\prime\prime}} + \beta^{}_{i^\prime}
- \gamma^{}_i\right)
\nonumber \\
&
- s^{}_{i4} s^{}_{i6} \left(s^2_{i5} - I^{-1}_{44} \left|I^{}_{45}\right|^2\right)
\sin 2\gamma^{}_i - s^{}_{i^\prime 4} s^{}_{i^\prime 6} \left(s^2_{i^\prime 5}
- I^{-1}_{44} \left|I^{}_{45}\right|^2\right) \sin\left(\gamma^{}_i +
\gamma^{}_{i^\prime}\right)
\nonumber \\
&
- s^{}_{i^{\prime\prime} 4} s^{}_{i^{\prime\prime} 6}
\left(s^2_{i^{\prime\prime} 5} - I^{-1}_{44} \left|I^{}_{45}\right|^2\right)
\sin\left(\gamma^{}_i + \gamma^{}_{i^{\prime\prime}}\right) \Big] \Bigg]
\nonumber \\
&
+ \Bigg[ \sum^3_{i=1} s^{}_{i 5} \Big[
s^{}_{i 6} \left(s^2_{i^\prime 4} + s^2_{i^{\prime\prime} 4}
\right) \cos\beta^{}_i - s^{}_{i4} s^{}_{i^\prime 4} s^{}_{i^\prime 6}
\cos\left(\alpha^{}_i + \gamma^{}_{i^\prime}\right) - s^{}_{i4}
s^{}_{i^{\prime\prime} 4} s^{}_{i^{\prime\prime} 6} \cos\left(\alpha^{}_i
+ \gamma^{}_{i^{\prime\prime}}\right)\Big]\Bigg]
\nonumber \\
&
\times \Bigg[
\sum^3_{i=1} s^2_{i4} s^{}_{i^\prime 5} s^{}_{i^{\prime\prime} 5}
\Big[s^{}_{i4} s^{}_{i5} s^{}_{i^\prime 4} s^{}_{i^{\prime\prime} 6}
\sin\left(\alpha^{}_i + \alpha^{}_{i^{\prime}} - \beta^{}_{i^{\prime\prime}}\right)
- s^{}_{i4} s^{}_{i5} s^{}_{i^\prime 6} s^{}_{i^{\prime\prime} 4}
\sin\left(\alpha^{}_i + \alpha^{}_{i^{\prime\prime}} - \beta^{}_{i^\prime}\right)
\nonumber \\
&
+ s^2_{i^\prime 4} s^{}_{i^\prime 5} s^{}_{i^{\prime\prime} 6}
\sin\left(2\alpha^{}_{i^\prime} - \beta^{}_{i^{\prime\prime}}\right)
- s^{}_{i^\prime 4} s^{}_{i^\prime 5} s^{}_{i^\prime 6} s^{}_{i^{\prime\prime} 4}
\sin\left(\alpha^{}_{i^{\prime}} + \alpha^{}_{i^{\prime\prime}} - \beta^{}_{i^{\prime}}
\right)
\nonumber \\
&
+ s^{}_{i^{\prime\prime} 4} s^{}_{i^{\prime\prime} 5} s^{}_{i^\prime 4}
s^{}_{i^{\prime\prime} 6} \sin\left(\alpha^{}_{i^\prime} + \alpha^{}_{i^{\prime\prime}}
- \beta^{}_{i^{\prime\prime}}\right) - s^{}_{i^\prime 6} s^2_{i^{\prime\prime} 4}
s^{}_{i^{\prime\prime} 5} \sin\left(2\alpha^{}_{i^{\prime\prime}} -
\beta^{}_{i^\prime}\right) \Big] \Bigg]
\nonumber \\
&
+ \Bigg[ \sum^3_{i=1} s^2_{i4} s^{}_{i^\prime 5} s^{}_{i^{\prime\prime} 5}
\Big[s^{}_{i4} s^{}_{i5} s^{}_{i^\prime 6} s^{}_{i^{\prime\prime} 4}
\cos\left(\alpha^{}_i + \alpha^{}_{i^{\prime\prime}} - \beta^{}_{i^{\prime}}\right)
- s^{}_{i4} s^{}_{i5} s^{}_{i^\prime 4} s^{}_{i^{\prime\prime} 6}
\cos\left(\alpha^{}_i + \alpha^{}_{i^{\prime}} - \beta^{}_{i^{\prime\prime}}\right)
\nonumber \\
&
+ s^{}_{i^\prime 4} s^{}_{i^\prime 5} s^{}_{i^\prime 6} s^{}_{i^{\prime\prime} 4}
\cos\left(\alpha^{}_{i^\prime} + \alpha^{}_{i^{\prime\prime}}
- \beta^{}_{i^{\prime}}\right)
- s^2_{i^\prime 4} s^{}_{i^\prime 5} s^{}_{i^{\prime\prime} 6}
\cos\left(2\alpha^{}_{i^{\prime}} - \beta^{}_{i^{\prime\prime}}\right)
\nonumber \\
&
+ s^{}_{i^{\prime} 6} s^2_{i^{\prime\prime} 4} s^{}_{i^{\prime\prime} 5}
\cos\left(2\alpha^{}_{i^{\prime\prime}} - \beta^{}_{i^{\prime}}\right)
- s^{}_{i^\prime 4} s^{}_{i^{\prime\prime} 4} s^{}_{i^{\prime\prime} 5}
s^{}_{i^{\prime\prime} 6} \cos\left(\alpha^{}_{i^\prime} + \alpha^{}_{i^{\prime\prime}}
- \beta^{}_{i^{\prime\prime}}\right) \Big] \Bigg]
\nonumber \\
&
\times \Bigg[ \sum^3_{i=1} s^{}_{i 5} \Big[
s^{}_{i 6} \left(s^2_{i^\prime 4} + s^2_{i^{\prime\prime} 4}\right)
\sin\beta^{}_i + s^{}_{i4} s^{}_{i^\prime 4} s^{}_{i^\prime 6}
\sin\left(\alpha^{}_i + \gamma^{}_{i^\prime}\right) + s^{}_{i4}
s^{}_{i^{\prime\prime} 4} s^{}_{i^{\prime\prime} 6} \sin\left(\alpha^{}_i
+ \gamma^{}_{i^{\prime\prime}}\right)\Big]\Bigg] \Bigg\}
\nonumber \\
&
+ {\rm term}\left\{\left(4, 5, 6\right) \to \left(4, 6, 5\right); \;
\left(\alpha^{}_i , \beta^{}_i , \gamma^{}_i\right) \to
-\left(\gamma^{}_i , \beta^{}_i , \alpha^{}_i\right)\right\}
\nonumber \\
&
+ {\rm term}\left\{\left(4, 5, 6\right) \to \left(5, 4, 6\right); \;
\left(\alpha^{}_i , \beta^{}_i , \gamma^{}_i\right) \to
-\left(\alpha^{}_i , \gamma^{}_i , \beta^{}_i\right)\right\}
\nonumber \\
&
+ {\rm term}\left\{\left(4, 5, 6\right) \to \left(5, 6, 4\right); \;
\left(\alpha^{}_i , \beta^{}_i , \gamma^{}_i\right) \to
+\left(\beta^{}_i , \gamma^{}_i , \alpha^{}_i\right)\right\}
\nonumber \\
&
+ {\rm term}\left\{\left(4, 5, 6\right) \to \left(6, 4, 5\right); \;
\left(\alpha^{}_i , \beta^{}_i , \gamma^{}_i\right) \to
+\left(\gamma^{}_i , \alpha^{}_i , \beta^{}_i\right)\right\}
\nonumber \\
&
+ {\rm term}\left\{\left(4, 5, 6\right) \to \left(6, 5, 4\right); \;
\left(\alpha^{}_i , \beta^{}_i , \gamma^{}_i\right) \to
-\left(\beta^{}_i , \alpha^{}_i , \gamma^{}_i\right)\right\} \; ,
\label{16}
\end{align}
where the subscripts $\left(i, i^\prime, i^{\prime\prime}\right)$ run
cyclically over $\left(1, 2, 3\right)$, and the {\it heavy} subscript replacements
for those abbreviation terms have the same implications as in Eq.~(\ref{15}).

\subsection{Term $T^{}_{\bf 222}$}

Finally, the terms of ${\rm Im}\left[(M^{}_\nu M^\dagger_\nu)^{}_{e\mu}
(M^{}_\nu M^\dagger_\nu)^{}_{\mu\tau} (M^{}_\nu M^\dagger_\nu)^{}_{\tau e}\right]$
that are proportional to $M^2_4 M^2_5 M^2_6$ are summarily denoted as $T^{}_{\bf 222}$.
The explicit analytical result of $T^{}_{\bf 222}$ is found to be
\begin{align}
T^{}_{\bf 222} = & \hspace{0.13cm}
M^2_4 M^2_5 M^2_6 \Bigg\{
I^{}_{44} I^{}_{55} I^{}_{66} s^{}_{14} s^{}_{15} s^{}_{26} s^{}_{36}
\Big[ s^{}_{25} s^{}_{34} \sin\left(\alpha^{}_1 + \beta^{}_2 + \gamma^{}_3\right)
- s^{}_{24} s^{}_{35} \sin\left(\alpha^{}_1 + \beta^{}_3 + \gamma^{}_2\right)\Big]
\nonumber \\
&
+ \left(I^{}_{55} \left|I^{}_{64}\right|^2 - I^{}_{44} \left|I^{}_{56}\right|^2\right)
\Bigg[ \sum^3_{i=1} s^2_{i6} s^{}_{i^\prime 4} s^{}_{i^\prime 5}
s^{}_{i^{\prime\prime} 4} s^{}_{i^{\prime\prime} 5} \sin\left(\alpha^{}_{i^\prime}
- \alpha^{}_{i^{\prime\prime}}\right) \Bigg]
\nonumber \\
&
- I^{}_{66} \Bigg[
\sum^3_{i=1} s^{}_{i4} s^{}_{i5} \Big[s^{}_{i4} s^{}_{i5} \cos 2\alpha^{}_i
+ 2 s^{}_{i^\prime 4} s^{}_{i^\prime 5} \cos\left(\alpha^{}_i + \alpha^{}_{i^\prime}\right)
\Big]\Bigg]
\nonumber \\
&
\times \Bigg[
\sum^3_{i=1} s^{}_{i4} s^{}_{i5} s^{}_{i^\prime 6} s^{}_{i^{\prime\prime} 6}
\Big[s^{}_{i^\prime 5} s^{}_{i^{\prime\prime} 4} \sin\left(\alpha^{}_i
- \beta^{}_{i^\prime} - \gamma^{}_{i^{\prime\prime}}\right)
- s^{}_{i^\prime 4} s^{}_{i^{\prime\prime} 5} \sin\left(\alpha^{}_i
- \beta^{}_{i^{\prime\prime}} - \gamma^{}_{i^{\prime}}\right)\Big]\Bigg]
\nonumber \\
&
- I^{}_{66} \Bigg[
\sum^3_{i=1} s^{}_{i4} s^{}_{i5} s^{}_{i^\prime 6} s^{}_{i^{\prime\prime} 6}
\Big[s^{}_{i^\prime 5} s^{}_{i^{\prime\prime} 4} \cos\left(\alpha^{}_i
- \beta^{}_{i^\prime} - \gamma^{}_{i^{\prime\prime}}\right)
- s^{}_{i^\prime 4} s^{}_{i^{\prime\prime} 5} \cos\left(\alpha^{}_i
- \beta^{}_{i^{\prime\prime}} - \gamma^{}_{i^{\prime}}\right)\Big]\Bigg]
\nonumber \\
&
\times \Bigg[
\sum^3_{i=1} s^{}_{i4} s^{}_{i5} \Big[s^{}_{i4} s^{}_{i5} \sin 2\alpha^{}_i
+ 2 s^{}_{i^\prime 4} s^{}_{i^\prime 5} \sin\left(\alpha^{}_i + \alpha^{}_{i^\prime}\right)
\Big]\Bigg]
\nonumber \\
&
+ 2 s^{}_{14} s^{}_{15} s^{}_{26} s^{}_{36} \Bigg[ \sum^3_{i=1} s^{}_{i4} s^{}_{i5} s^2_{i6}
\cos\left(\alpha^{}_1 + \alpha^{}_i\right)
+ \sum^3_{i=1} s^{}_{i6} s^{}_{i^\prime 6} \Big[s^{}_{i4} s^{}_{i^\prime 5}
\cos\left(\alpha^{}_1 - \beta^{}_{i^\prime} - \gamma^{}_i\right)
\nonumber \\
&
+ s^{}_{i5} s^{}_{i^\prime 4} \cos\left(\alpha^{}_1 - \beta^{}_i - \gamma^{}_{i^\prime}\right)
\Big]\Bigg]
\nonumber \\
&
\times \Bigg[ s^{}_{25} s^{}_{34} \sum^3_{i=1} s^{}_{i4} s^{}_{i5}
\sin\left(\alpha^{}_i - \beta^{}_2 - \gamma^{}_3\right)
- s^{}_{24} s^{}_{35} \sum^3_{i=1} s^{}_{i4} s^{}_{i5}
\sin\left(\alpha^{}_i - \beta^{}_3 - \gamma^{}_2\right) \Bigg]
\nonumber \\
&
- 2 s^{}_{14} s^{}_{15} s^{}_{26} s^{}_{36} \Bigg[
s^{}_{25} s^{}_{34} \sum^3_{i=1} s^{}_{i4} s^{}_{i6} \Big[
s^{}_{i5} s^{}_{i6} \cos\left(\beta^{}_i + \beta^{}_2 - \gamma^{}_i - \gamma^{}_3\right)
+ s^{}_{i^\prime 5} s^{}_{i^\prime 6}
\cos\left(\beta^{}_{i^\prime} + \beta^{}_2 - \gamma^{}_i - \gamma^{}_3\right)
\nonumber \\
&
+ s^{}_{i^{\prime\prime} 5} s^{}_{i^{\prime\prime} 6}
\cos\left(\beta^{}_{i^{\prime\prime}} + \beta^{}_2 - \gamma^{}_i - \gamma^{}_3\right)\Big]
- s^{}_{24} s^{}_{35} \sum^3_{i=1} s^{}_{i4} s^{}_{i6} \Big[
s^{}_{i5} s^{}_{i6} \cos\left(\beta^{}_i + \beta^{}_3 - \gamma^{}_i - \gamma^{}_2\right)
\nonumber \\
&
+ s^{}_{i^\prime 5} s^{}_{i^\prime 6}
\cos\left(\beta^{}_{i^\prime} + \beta^{}_3 - \gamma^{}_i - \gamma^{}_2\right)
+ s^{}_{i^{\prime\prime} 5} s^{}_{i^{\prime\prime} 6}
\cos\left(\beta^{}_{i^{\prime\prime}} + \beta^{}_3 - \gamma^{}_i - \gamma^{}_2\right)
\Big] \Bigg]
\nonumber \\
&
\times \Bigg[\sum^3_{i=1} s^{}_{i4} s^{}_{i5} \sin\left(\alpha^{}_1 + \alpha^{}_i\right)
\Bigg]
\nonumber \\
&
+ \left(s^2_{14} s^2_{25} - s^2_{15} s^2_{24}\right) s^2_{36}
\Bigg[ \sum^3_{i=1} \Big[ s^{}_{i4} s^{}_{i5} s^{}_{i^\prime 4} s^{}_{i^\prime 5}
\left(s^2_{i6} - s^2_{i^\prime 6}\right) \sin\left(\alpha^{}_i - \alpha^{}_{i^\prime}\right)
+ s^{}_{i5} s^{}_{i6} s^{}_{i^\prime 5} s^{}_{i^\prime 6}
\left(s^2_{i4} - s^2_{i^\prime 4}\right)
\nonumber \\
&
\times \sin\left(\beta^{}_i - \beta^{}_{i^\prime}\right)
+ s^{}_{i4} s^{}_{i6} s^{}_{i^\prime 4} s^{}_{i^\prime 6}
\left(s^2_{i5} - s^2_{i^\prime 5}\right) \sin\left(\gamma^{}_i - \gamma^{}_{i^\prime}\right)
\Big]
\nonumber \\
&
- \sum^3_{i=1} s^{}_{i4} s^{}_{i5} \Big[
s^{}_{i^\prime 5} s^{}_{i^\prime 6} s^{}_{i^{\prime\prime} 4} s^{}_{i^{\prime\prime} 6}
\sin\left(\alpha^{}_i + \beta^{}_{i^\prime} + \gamma^{}_{i^{\prime\prime}}\right)
+ s^{}_{i^\prime 4} s^{}_{i^\prime 6} s^{}_{i^{\prime\prime} 5} s^{}_{i^{\prime\prime} 6}
\sin\left(\alpha^{}_i + \beta^{}_{i^{\prime\prime}} + \gamma^{}_{i^{\prime}}\right)
\Big] \Bigg] \Bigg\}
\nonumber \\
&
+ {\rm term}\left\{\left(4, 5, 6\right) \to \left(4, 6, 5\right); \;
\left(\alpha^{}_i , \beta^{}_i , \gamma^{}_i\right) \to
-\left(\gamma^{}_i , \beta^{}_i , \alpha^{}_i\right)\right\}
\nonumber \\
&
+ {\rm term}\left\{\left(4, 5, 6\right) \to \left(6, 5, 4\right); \;
\left(\alpha^{}_i , \beta^{}_i , \gamma^{}_i\right) \to
-\left(\beta^{}_i , \alpha^{}_i , \gamma^{}_i\right)\right\} ,
\label{17}
\end{align}
where the three {\it light} subscripts $\left(i, i^\prime, i^{\prime\prime}\right)$ run
cyclically over $\left(1, 2, 3\right)$, and the {\it heavy} subscript replacements
for those abbreviation terms have the same implications as in Eq.~(\ref{14}).
It is straightforward to see that $T^{}_{\bf 411}$, $T^{}_{\bf 321}$ and $T^{}_{\bf 222}$
characterize the three-species interference effects in which all the three kinds of
heavy Majorana neutrinos take part, and hence their analytical expressions are much more
complicated than those of $T^{}_{\bf 33}$ and $T^{}_{\bf 42}$.

To summarize, a general and explicit expression of the Jarlskog invariant
${\cal J}^{}_\nu$ in the canonical seesaw mechanism turns out to be
\begin{align}
{\cal J}^{}_\nu = \frac{T^{}_{\bf 33} + T^{}_{\bf 42} + T^{}_{\bf 411} +
T^{}_{\bf 321} + T^{}_{\bf 222}}
{\Delta^{}_{21} \Delta^{}_{31} \Delta^{}_{32}} \; ,
\label{18}
\end{align}
where the analytical results of $T^{}_{\bf 33}$, $T^{}_{\bf 42}$, $T^{}_{\bf 411}$,
$T^{}_{\bf 321}$ and $T^{}_{\bf 222}$ have been given in Eqs.~(\ref{13})---(\ref{17}).

Let us emphasize that the analytical results obtained above are completely new,
and they provide the first general and explicit expression of ${\cal J}^{}_\nu$ in
terms of the original flavor parameters in the canonical seesaw mechanism.
It is certainly too lengthy to write out all the abbreviation terms in the above
equations, but one may figure out all the combinations of the CP-violating
phases $\alpha^{}_i$, $\beta^{}_i$ and $\gamma^{}_i$ (for $i = 1, 2, 3$) that
appear in the expression of ${\cal J}^{}_\nu$.

\subsection{Phase combinations}

After a very tedious survey of all the possibilities, including a careful count of
the phase combinations in those abbreviation terms, we find that there are totally
240 linear combinations of the 6 original seesaw phase parameters in ${\cal J}^{}_\nu$:
\begin{equation}
\begin{array}{lllllllll}
\sin\alpha^{}_1 , \;\; & \sin\alpha^{}_2 , \;\; & \sin\alpha^{}_3 ; \;\;
& \sin\beta^{}_1 , \;\; & \sin\beta^{}_2 , \;\; & \sin\beta^{}_3 ; \;\;
& \sin\gamma^{}_1 , \;\; & \sin\gamma^{}_2 , \;\; & \sin\gamma^{}_3 ;
\nonumber \\ \vspace{-0.4cm} \nonumber \\
\sin 2\alpha^{}_1 , & \sin 2\alpha^{}_2 , & \sin 2\alpha^{}_3 ;
& \sin 2\beta^{}_1 , & \sin 2\beta^{}_2 , & \sin 2\beta^{}_3 ;
& \sin 2\gamma^{}_1 , & \sin 2\gamma^{}_2 , & \sin 2\gamma^{}_3 ;
\nonumber
\end{array}
\end{equation}
\vspace{-0.25cm}
\begin{equation}
\begin{array}{llllll}
\sin\left(\alpha^{}_1 + \alpha^{}_2\right) ,
& \sin\left(\alpha^{}_2 + \alpha^{}_3\right) ,
& \sin\left(\alpha^{}_3 + \alpha^{}_1\right) ;
& \sin\left(\beta^{}_1 + \beta^{}_2\right) ,
& \sin\left(\beta^{}_2 + \beta^{}_3\right) ,
& \sin\left(\beta^{}_3 + \beta^{}_1\right) ; \hspace{1.9cm}
\nonumber \\
\sin\left(\gamma^{}_1 + \gamma^{}_2\right) ,
& \sin\left(\gamma^{}_2 + \gamma^{}_3\right) ,
& \sin\left(\gamma^{}_3 + \gamma^{}_1\right) ;
&&&
\nonumber \\
\sin\left(\alpha^{}_1 - \alpha^{}_2\right) ,
& \sin\left(\alpha^{}_2 - \alpha^{}_3\right) ,
& \sin\left(\alpha^{}_3 - \alpha^{}_1\right) ;
& \sin\left(\beta^{}_1 - \beta^{}_2\right) ,
& \sin\left(\beta^{}_2 - \beta^{}_3\right) ,
& \sin\left(\beta^{}_3 - \beta^{}_1\right) ; \hspace{1.9cm}
\nonumber \\
\sin\left(\gamma^{}_1 - \gamma^{}_2\right) ,
& \sin\left(\gamma^{}_2 - \gamma^{}_3\right) ,
& \sin\left(\gamma^{}_3 - \gamma^{}_1\right) ;
&&&
\nonumber \\ \vspace{-0.3cm} \nonumber \\
\sin\left(\alpha^{}_1 + \beta^{}_2\right) ,
& \sin\left(\alpha^{}_1 + \beta^{}_3\right) ,
& \sin\left(\alpha^{}_2 + \beta^{}_1\right) ,
& \sin\left(\alpha^{}_2 + \beta^{}_3\right) ,
& \sin\left(\alpha^{}_3 + \beta^{}_1\right) ,
& \sin\left(\alpha^{}_3 + \beta^{}_2\right) ;
\nonumber \\
\sin\left(\alpha^{}_1 + \gamma^{}_2\right) ,
& \sin\left(\alpha^{}_1 + \gamma^{}_3\right) ,
& \sin\left(\alpha^{}_2 + \gamma^{}_1\right) ,
& \sin\left(\alpha^{}_2 + \gamma^{}_3\right) ,
& \sin\left(\alpha^{}_3 + \gamma^{}_1\right) ,
& \sin\left(\alpha^{}_3 + \gamma^{}_2\right) ;
\nonumber \\
\sin\left(\beta^{}_1 + \gamma^{}_2\right) ,
& \sin\left(\beta^{}_1 + \gamma^{}_3\right) ,
& \sin\left(\beta^{}_2 + \gamma^{}_1\right) ,
& \sin\left(\beta^{}_2 + \gamma^{}_3\right) ,
& \sin\left(\beta^{}_3 + \gamma^{}_1\right) ,
& \sin\left(\beta^{}_3 + \gamma^{}_2\right) ; \hspace{1.6cm}
\nonumber \\ \vspace{-0.3cm} \nonumber \\
\sin 2\left(\alpha^{}_1 + \beta^{}_2\right) ,
& \sin 2\left(\alpha^{}_1 + \beta^{}_3\right) ,
& \sin 2\left(\alpha^{}_2 + \beta^{}_1\right) ,
& \sin 2\left(\alpha^{}_2 + \beta^{}_3\right) ,
& \sin 2\left(\alpha^{}_3 + \beta^{}_1\right) ,
& \sin 2\left(\alpha^{}_3 + \beta^{}_2\right) ;
\nonumber \\
\sin 2\left(\alpha^{}_1 + \gamma^{}_2\right) ,
& \sin 2\left(\alpha^{}_1 + \gamma^{}_3\right) ,
& \sin 2\left(\alpha^{}_2 + \gamma^{}_1\right) ,
& \sin 2\left(\alpha^{}_2 + \gamma^{}_3\right) ,
& \sin 2\left(\alpha^{}_3 + \gamma^{}_1\right) ,
& \sin 2\left(\alpha^{}_3 + \gamma^{}_2\right) ;
\nonumber \\
\sin 2\left(\beta^{}_1 + \gamma^{}_2\right) ,
& \sin 2\left(\beta^{}_1 + \gamma^{}_3\right) ,
& \sin 2\left(\beta^{}_2 + \gamma^{}_1\right) ,
& \sin 2\left(\beta^{}_2 + \gamma^{}_3\right) ,
& \sin 2\left(\beta^{}_3 + \gamma^{}_1\right) ,
& \sin 2\left(\beta^{}_3 + \gamma^{}_2\right) ;
\nonumber \\ \vspace{-0.3cm}  \nonumber \\
\sin\left(2\alpha^{}_1 - \beta^{}_2\right) ,
& \sin\left(2\alpha^{}_1 - \beta^{}_3\right) ,
& \sin\left(2\alpha^{}_2 - \beta^{}_1\right) ,
& \sin\left(2\alpha^{}_2 - \beta^{}_3\right) ,
& \sin\left(2\alpha^{}_3 - \beta^{}_1\right) ,
& \sin\left(2\alpha^{}_3 - \beta^{}_2\right) ;
\nonumber \\
\sin\left(2\alpha^{}_1 - \gamma^{}_2\right) ,
& \sin\left(2\alpha^{}_1 - \gamma^{}_3\right) ,
& \sin\left(2\alpha^{}_2 - \gamma^{}_1\right) ,
& \sin\left(2\alpha^{}_2 - \gamma^{}_3\right) ,
& \sin\left(2\alpha^{}_3 - \gamma^{}_1\right) ,
& \sin\left(2\alpha^{}_3 - \gamma^{}_2\right) ;
\nonumber \\
\sin\left(2\beta^{}_1 - \alpha^{}_2\right) ,
& \sin\left(2\beta^{}_1 - \alpha^{}_3\right) ,
& \sin\left(2\beta^{}_2 - \alpha^{}_1\right) ,
& \sin\left(2\beta^{}_2 - \alpha^{}_3\right) ,
& \sin\left(2\beta^{}_3 - \alpha^{}_1\right) ,
& \sin\left(2\beta^{}_3 - \alpha^{}_2\right) ;
\nonumber \\
\sin\left(2\beta^{}_1 - \gamma^{}_2\right) ,
& \sin\left(2\beta^{}_1 - \gamma^{}_3\right) ,
& \sin\left(2\beta^{}_2 - \gamma^{}_1\right) ,
& \sin\left(2\beta^{}_2 - \gamma^{}_3\right) ,
& \sin\left(2\beta^{}_3 - \gamma^{}_1\right) ,
& \sin\left(2\beta^{}_3 - \gamma^{}_2\right) ;
\nonumber \\
\sin\left(2\gamma^{}_1 - \alpha^{}_2\right) ,
& \sin\left(2\gamma^{}_1 - \alpha^{}_3\right) ,
& \sin\left(2\gamma^{}_2 - \alpha^{}_1\right) ,
& \sin\left(2\gamma^{}_2 - \alpha^{}_3\right) ,
& \sin\left(2\gamma^{}_3 - \alpha^{}_1\right) ,
& \sin\left(2\gamma^{}_3 - \alpha^{}_2\right) ;
\nonumber \\
\sin\left(2\gamma^{}_1 - \beta^{}_2\right) ,
& \sin\left(2\gamma^{}_1 - \beta^{}_3\right) ,
& \sin\left(2\gamma^{}_2 - \beta^{}_1\right) ,
& \sin\left(2\gamma^{}_2 - \beta^{}_3\right) ,
& \sin\left(2\gamma^{}_3 - \beta^{}_1\right) ,
& \sin\left(2\gamma^{}_3 - \beta^{}_2\right) ; \hspace{0.35cm}
\nonumber
\end{array}
\end{equation}
\begin{equation}
\begin{array}{llll}
\sin\left(\alpha^{}_1 + \alpha^{}_2 + 2\beta^{}_3\right) ,
& \sin\left(\alpha^{}_1 + \alpha^{}_2 + 2\gamma^{}_3\right) ,
& \sin\left(\alpha^{}_1 + \alpha^{}_3 + 2\beta^{}_2\right) ,
& \sin\left(\alpha^{}_1 + \alpha^{}_3 + 2\gamma^{}_2\right) ,
\nonumber \\
\sin\left(\alpha^{}_2 + \alpha^{}_3 + 2\beta^{}_1\right) ,
& \sin\left(\alpha^{}_2 + \alpha^{}_3 + 2\gamma^{}_1\right) ;
& \sin\left(\beta^{}_1 + \beta^{}_2 + 2\alpha^{}_3\right) ,
& \sin\left(\beta^{}_1 + \beta^{}_2 + 2\gamma^{}_3\right) ,
\nonumber \\
\sin\left(\beta^{}_1 + \beta^{}_3 + 2\alpha^{}_2\right) ,
& \sin\left(\beta^{}_1 + \beta^{}_3 + 2\gamma^{}_2\right) ,
& \sin\left(\beta^{}_2 + \beta^{}_3 + 2\alpha^{}_1\right) ,
& \sin\left(\beta^{}_2 + \beta^{}_3 + 2\gamma^{}_1\right) ;
\nonumber\\
\sin\left(\gamma^{}_1 + \gamma^{}_2 + 2\alpha^{}_3\right) ,
& \sin\left(\gamma^{}_1 + \gamma^{}_2 + 2\beta^{}_3\right) ,
& \sin\left(\gamma^{}_1 + \gamma^{}_3 + 2\alpha^{}_2\right) ,
& \sin\left(\gamma^{}_1 + \gamma^{}_3 + 2\beta^{}_2\right) ,
\nonumber \\
\sin\left(\gamma^{}_2 + \gamma^{}_3 + 2\alpha^{}_1\right) ,
& \sin\left(\gamma^{}_2 + \gamma^{}_3 + 2\beta^{}_1\right) ;
&&
\nonumber \\ \vspace{-0.3cm} \nonumber \\
\sin\left(\alpha^{}_1 + \beta^{}_2 + \gamma^{}_3\right) ,
& \sin\left(\alpha^{}_1 + \beta^{}_3 + \gamma^{}_2\right) ,
& \sin\left(\alpha^{}_2 + \beta^{}_1 + \gamma^{}_3\right) ,
& \sin\left(\alpha^{}_2 + \beta^{}_3 + \gamma^{}_1\right) ,
\nonumber \\
\sin\left(\alpha^{}_3 + \beta^{}_1 + \gamma^{}_2\right) ,
& \sin\left(\alpha^{}_3 + \beta^{}_2 + \gamma^{}_1\right) .
&&
\nonumber
\end{array}
\end{equation}
\begin{equation}
\begin{array}{llll}
\sin\left(\alpha^{}_1 + \alpha^{}_2 - \beta^{}_1\right) ,
& \sin\left(\alpha^{}_1 + \alpha^{}_2 - \beta^{}_2\right) ,
& \sin\left(\alpha^{}_1 + \alpha^{}_2 - \beta^{}_3\right) ,
& \sin\left(\alpha^{}_1 + \alpha^{}_3 - \beta^{}_1\right) ,
\nonumber \\
\sin\left(\alpha^{}_1 + \alpha^{}_3 - \beta^{}_2\right) ,
& \sin\left(\alpha^{}_1 + \alpha^{}_3 - \beta^{}_3\right) ,
& \sin\left(\alpha^{}_2 + \alpha^{}_3 - \beta^{}_1\right) ,
& \sin\left(\alpha^{}_2 + \alpha^{}_3 - \beta^{}_2\right) ,
\nonumber \\
\sin\left(\alpha^{}_2 + \alpha^{}_3 - \beta^{}_3\right) ;
& \sin\left(\alpha^{}_1 + \alpha^{}_2 - \gamma^{}_1\right) ,
& \sin\left(\alpha^{}_1 + \alpha^{}_2 - \gamma^{}_2\right) ,
& \sin\left(\alpha^{}_1 + \alpha^{}_2 - \gamma^{}_3\right) ,
\nonumber \\
\sin\left(\alpha^{}_1 + \alpha^{}_3 - \gamma^{}_1\right) ,
& \sin\left(\alpha^{}_1 + \alpha^{}_3 - \gamma^{}_2\right) ,
& \sin\left(\alpha^{}_1 + \alpha^{}_3 - \gamma^{}_3\right) ,
& \sin\left(\alpha^{}_2 + \alpha^{}_3 - \gamma^{}_1\right) ,
\nonumber \\
\sin\left(\alpha^{}_2 + \alpha^{}_3 - \gamma^{}_2\right) ,
& \sin\left(\alpha^{}_2 + \alpha^{}_3 - \gamma^{}_3\right) ;
& \sin\left(\beta^{}_1 + \beta^{}_2 - \alpha^{}_1\right) ,
& \sin\left(\beta^{}_1 + \beta^{}_2 - \alpha^{}_2\right) ,
\nonumber \\
\sin\left(\beta^{}_1 + \beta^{}_2 - \alpha^{}_3\right) ,
& \sin\left(\beta^{}_1 + \beta^{}_3 - \alpha^{}_1\right) ,
& \sin\left(\beta^{}_1 + \beta^{}_3 - \alpha^{}_2\right) ,
& \sin\left(\beta^{}_1 + \beta^{}_3 - \alpha^{}_3\right) ,
\nonumber \\
\sin\left(\beta^{}_2 + \beta^{}_3 - \alpha^{}_1\right) ,
& \sin\left(\beta^{}_2 + \beta^{}_3 - \alpha^{}_2\right) ,
& \sin\left(\beta^{}_2 + \beta^{}_3 - \alpha^{}_3\right) ;
& \sin\left(\beta^{}_1 + \beta^{}_2 - \gamma^{}_1\right) ,
\nonumber \\
\sin\left(\beta^{}_1 + \beta^{}_2 - \gamma^{}_2\right) ,
& \sin\left(\beta^{}_1 + \beta^{}_2 - \gamma^{}_3\right) ,
& \sin\left(\beta^{}_1 + \beta^{}_3 - \gamma^{}_1\right) ,
& \sin\left(\beta^{}_1 + \beta^{}_3 - \gamma^{}_2\right) ,
\nonumber \\
\sin\left(\beta^{}_1 + \beta^{}_3 - \gamma^{}_3\right) ,
& \sin\left(\beta^{}_2 + \beta^{}_3 - \gamma^{}_1\right) ,
& \sin\left(\beta^{}_2 + \beta^{}_3 - \gamma^{}_2\right) ,
& \sin\left(\beta^{}_2 + \beta^{}_3 - \gamma^{}_3\right) ;
\nonumber \\
\sin\left(\gamma^{}_1 + \gamma^{}_2 - \alpha^{}_1\right) ,
& \sin\left(\gamma^{}_1 + \gamma^{}_2 - \alpha^{}_2\right) ,
& \sin\left(\gamma^{}_1 + \gamma^{}_2 - \alpha^{}_3\right) ,
& \sin\left(\gamma^{}_1 + \gamma^{}_3 - \alpha^{}_1\right) ,
\nonumber \\
\sin\left(\gamma^{}_1 + \gamma^{}_3 - \alpha^{}_2\right) ,
& \sin\left(\gamma^{}_1 + \gamma^{}_3 - \alpha^{}_3\right) ,
& \sin\left(\gamma^{}_2 + \gamma^{}_3 - \alpha^{}_1\right) ,
& \sin\left(\gamma^{}_2 + \gamma^{}_3 - \alpha^{}_2\right) ,
\nonumber \\
\sin\left(\gamma^{}_2 + \gamma^{}_3 - \alpha^{}_3\right) ;
& \sin\left(\gamma^{}_1 + \gamma^{}_2 - \beta^{}_1\right) ,
& \sin\left(\gamma^{}_1 + \gamma^{}_2 - \beta^{}_2\right) ,
& \sin\left(\gamma^{}_1 + \gamma^{}_2 - \beta^{}_3\right) ,
\nonumber \\
\sin\left(\gamma^{}_1 + \gamma^{}_3 - \beta^{}_1\right) ,
& \sin\left(\gamma^{}_1 + \gamma^{}_3 - \beta^{}_2\right) ,
& \sin\left(\gamma^{}_1 + \gamma^{}_3 - \beta^{}_3\right) ,
& \sin\left(\gamma^{}_2 + \gamma^{}_3 - \beta^{}_1\right) ,
\nonumber \\
\sin\left(\gamma^{}_2 + \gamma^{}_3 - \beta^{}_2\right) ,
& \sin\left(\gamma^{}_2 + \gamma^{}_3 - \beta^{}_3\right) ;
&&
\nonumber \\ \vspace{-0.3cm} \nonumber \\
\sin\left(\alpha^{}_1 + \beta^{}_2 - \gamma^{}_1\right) ,
& \sin\left(\alpha^{}_1 + \beta^{}_2 - \gamma^{}_2\right) ,
& \sin\left(\alpha^{}_1 + \beta^{}_2 - \gamma^{}_3\right) ,
& \sin\left(\alpha^{}_1 + \beta^{}_3 - \gamma^{}_1\right) ,
\nonumber \\
\sin\left(\alpha^{}_1 + \beta^{}_3 - \gamma^{}_2\right) ,
& \sin\left(\alpha^{}_1 + \beta^{}_3 - \gamma^{}_3\right) ,
& \sin\left(\alpha^{}_2 + \beta^{}_1 - \gamma^{}_1\right) ,
& \sin\left(\alpha^{}_2 + \beta^{}_1 - \gamma^{}_2\right) ,
\nonumber \\
\sin\left(\alpha^{}_2 + \beta^{}_1 - \gamma^{}_3\right) ,
& \sin\left(\alpha^{}_2 + \beta^{}_3 - \gamma^{}_1\right) ,
& \sin\left(\alpha^{}_2 + \beta^{}_3 - \gamma^{}_2\right) ,
& \sin\left(\alpha^{}_2 + \beta^{}_3 - \gamma^{}_3\right) ,
\nonumber \\
\sin\left(\alpha^{}_3 + \beta^{}_1 - \gamma^{}_1\right) ,
& \sin\left(\alpha^{}_3 + \beta^{}_1 - \gamma^{}_2\right) ,
& \sin\left(\alpha^{}_3 + \beta^{}_1 - \gamma^{}_3\right) ,
& \sin\left(\alpha^{}_3 + \beta^{}_2 - \gamma^{}_1\right) ,
\nonumber \\
\sin\left(\alpha^{}_3 + \beta^{}_2 - \gamma^{}_2\right) ,
& \sin\left(\alpha^{}_3 + \beta^{}_2 - \gamma^{}_3\right) ;
& \sin\left(\alpha^{}_1 - \beta^{}_1 + \gamma^{}_2\right) ,
& \sin\left(\alpha^{}_1 - \beta^{}_1 + \gamma^{}_3\right) ,
\nonumber \\
\sin\left(\alpha^{}_1 - \beta^{}_2 + \gamma^{}_2\right) ,
& \sin\left(\alpha^{}_1 - \beta^{}_2 + \gamma^{}_3\right) ,
& \sin\left(\alpha^{}_1 - \beta^{}_3 + \gamma^{}_2\right) ,
& \sin\left(\alpha^{}_1 - \beta^{}_3 + \gamma^{}_3\right) ,
\nonumber \\
\sin\left(\alpha^{}_2 - \beta^{}_1 + \gamma^{}_1\right) ,
& \sin\left(\alpha^{}_2 - \beta^{}_1 + \gamma^{}_3\right) ,
& \sin\left(\alpha^{}_2 - \beta^{}_2 + \gamma^{}_1\right) ,
& \sin\left(\alpha^{}_2 - \beta^{}_2 + \gamma^{}_3\right) ,
\nonumber \\
\sin\left(\alpha^{}_2 - \beta^{}_3 + \gamma^{}_1\right) ,
& \sin\left(\alpha^{}_2 - \beta^{}_3 + \gamma^{}_3\right) ,
& \sin\left(\alpha^{}_3 - \beta^{}_1 + \gamma^{}_1\right) ,
& \sin\left(\alpha^{}_3 - \beta^{}_1 + \gamma^{}_2\right) ,
\nonumber \\
\sin\left(\alpha^{}_3 - \beta^{}_2 + \gamma^{}_1\right) ,
& \sin\left(\alpha^{}_3 - \beta^{}_2 + \gamma^{}_2\right) ,
& \sin\left(\alpha^{}_3 - \beta^{}_3 + \gamma^{}_1\right) ,
& \sin\left(\alpha^{}_3 - \beta^{}_3 + \gamma^{}_2\right) ;
\nonumber \\
\sin\left(\alpha^{}_1 - \beta^{}_1 - \gamma^{}_2\right) ,
& \sin\left(\alpha^{}_1 - \beta^{}_1 - \gamma^{}_3\right) ,
& \sin\left(\alpha^{}_1 - \beta^{}_2 - \gamma^{}_1\right) ,
& \sin\left(\alpha^{}_1 - \beta^{}_2 - \gamma^{}_3\right) ,
\nonumber \\
\sin\left(\alpha^{}_1 - \beta^{}_3 - \gamma^{}_1\right) ,
& \sin\left(\alpha^{}_1 - \beta^{}_3 - \gamma^{}_2\right) ,
& \sin\left(\alpha^{}_2 - \beta^{}_1 - \gamma^{}_2\right) ,
& \sin\left(\alpha^{}_2 - \beta^{}_1 - \gamma^{}_3\right) ,
\nonumber \\
\sin\left(\alpha^{}_2 - \beta^{}_2 - \gamma^{}_1\right) ,
& \sin\left(\alpha^{}_2 - \beta^{}_2 - \gamma^{}_3\right) ,
& \sin\left(\alpha^{}_2 - \beta^{}_3 - \gamma^{}_1\right) ,
& \sin\left(\alpha^{}_2 - \beta^{}_3 - \gamma^{}_2\right) ,
\nonumber \\
\sin\left(\alpha^{}_3 - \beta^{}_1 - \gamma^{}_2\right) ,
& \sin\left(\alpha^{}_3 - \beta^{}_1 - \gamma^{}_3\right) ,
& \sin\left(\alpha^{}_3 - \beta^{}_2 - \gamma^{}_1\right) ,
& \sin\left(\alpha^{}_3 - \beta^{}_2 - \gamma^{}_3\right) ,
\nonumber \\
\sin\left(\alpha^{}_3 - \beta^{}_3 - \gamma^{}_1\right) ,
& \sin\left(\alpha^{}_3 - \beta^{}_3 - \gamma^{}_2\right) .
&&
\nonumber
\end{array}
\end{equation}
Note that the number of such linear phase combinations will not be reduced
if $\gamma^{}_i = - \left(\alpha^{}_i + \beta^{}_i\right)$ is taken
into account. That is why we have kept $\gamma^{}_1$, $\gamma^{}_2$ and $\gamma^{}_3$
in the above list as in Eqs.~(\ref{13})---(\ref{17}), just for the sake of simplicity.
But one should keep in mind that only $\alpha^{}_i$ and $\beta^{}_i$ are the
independent phase parameters chosen in this work.

Of course, it is always possible to simplify the above linear phase combinations
to the forms of $\sin\alpha^{}_i$ and $\sin\beta^{}_i$ (for $i = 1, 2, 3$)
multiplied by the corresponding coefficients, such that the Jarlskog invariant
${\cal J}^{}_\nu$ can be compactly expressed as
\begin{align}
{\cal J}^{}_\nu = \sum^3_{i=1} \left(C^{}_{\alpha i} \sin\alpha^{}_i +
C^{}_{\beta i} \sin\beta^{}_i\right) \; ,
\label{19}
\end{align}
where the coefficients $C^{}_{\alpha i}$ and $C^{}_{\beta i}$ (for $i = 1, 2, 3$)
are some formidably complicated algebraic functions of the original seesaw flavor
parameters, including $\cos\alpha^{}_i$ and $\cos\beta^{}_i$ (for $i = 1, 2, 3$).
In this way one may actually extract the explicit expressions of $C^{}_{\alpha i}$
and $C^{}_{\beta i}$ from Eqs.~(\ref{13})---(\ref{17}), but the relevant results
are too lengthy to be presented here.

It is worth pointing out that the 240 phase combinations listed above will be greatly
reduced to~\cite{Xing:2023kdj}
\begin{align}
\sin 2\alpha^{}_1 \; , \;\; \sin 2\alpha^{}_2 \; , \;\; \sin 2\alpha^{}_3 \; ; \;\;
\sin\left(\alpha^{}_{1} \pm \alpha^{}_2\right) \; , \;\;
\sin\left(\alpha^{}_{2} \pm \alpha^{}_3\right) \; , \;\;
\sin\left(\alpha^{}_{3} \pm \alpha^{}_1\right) \; ,
\label{20}
\end{align}
if the third heavy Majorana neutrino species $N^{}_6$ is switched off, in which case
the canonical seesaw mechanism is simplified to the so-called minimal seesaw
scenario~\cite{Kleppe:1995zz,Ma:1998zg,Frampton:2002qc,Xing:2020ald}
\footnote{In Ref.~\cite{Fujihara:2005pv} the authors have successfully derived the
exact correlation between the Jarlskog invariant and the CP-violating asymmetries
of heavy Majorana neutrino decays by using a different parametrization of
the flavor structure for the minimal seesaw mechanism, although their results are
not as intuitional as the approximate ones obtained in the Euler-like block
parametrization of the minimal seesaw scenario~\cite{Xing:2023kdj}. It is extremely complicated, however, to do the same thing in the canonical seesaw framework
containing three heavy Majorana neutrino fields. The present work makes the
{\it first} attempt in this regard, but a compromised balance has been struck
between calculability of the observable quantities and complexity of their
analytical expressions (i.e., we have adopted the approximate expressions
of $A^{-1} R$ in our calculations from the very beginning). Even in this
situation our analytical results have clearly shown how complicated the
canonical (3+3) flavor case is, as compared with the minimal (3+2) flavor case.}.
Such an amazing simplification of the expression of ${\cal J}^{}_\nu$ implies that
a lot of information about CP violation in the canonical case is lost in the
minimal seesaw framework, although the latter is a good benchmark model.

\section{Comments on $\varepsilon^{}_{j\alpha}$}

Now let us examine how CP violation in the lepton-number-violating decays of
three heavy Majorana neutrinos depends on the original seesaw phase parameters.
Far above the electroweak symmetry breaking scale, one may calculate
the CP-violating asymmetries between $N^{}_j$ decays into the leptonic doublet
$\ell^{}_\alpha$ plus the Higgs doublet $H$ and their CP-conjugated processes
(for $j = 4, 5, 6$ and $\alpha = e, \mu, \tau$) at the one-loop level in
the canonical seesaw mechanism~\cite{Luty:1992un,Covi:1996wh,Plumacher:1996kc,
Pilaftsis:1997jf,Buchmuller:2005eh,DiBari:2021fhs}. With the help of our Euler-like
block parametrization of the seesaw flavor structure, we explicitly obtain
\begin{align}
\varepsilon^{}_{j \alpha} \equiv & \hspace{0.13cm}
\frac{\Gamma({N}^{}_j \to \ell^{}_\alpha + H)
- \Gamma({N}^{}_j \to \overline{\ell^{}_\alpha} +
\overline{H})}{\displaystyle \sum_\alpha \left[\Gamma({N}^{}_j \to
\ell^{}_\alpha + H) + \Gamma({N}^{}_j \to \overline{\ell^{}_\alpha}
+ \overline{H})\right]}
\nonumber \\
\simeq & \hspace{0.13cm}
\frac{1}{\displaystyle 8\pi \langle H\rangle^2 \sum_\beta \left|R^{}_{\beta j}\right|^2}
\sum^6_{k=4} \Bigg\{ M^2_k \hspace{0.1cm} {\rm Im} \Bigg[
\left(R^*_{\alpha j} R^{}_{\alpha k}\right) \sum_\beta \Big[ \left(R^*_{\beta j}
R^{}_{\beta k}\right) \xi(x^{}_{kj}) + \left(R^{}_{\beta j}
R^*_{\beta k}\right) \zeta(x^{}_{kj})\Big] \Bigg] \Bigg\} \; ,
\label{21}
\end{align}
where the Latin and Greek subscripts run respectively over $(4, 5, 6)$ and
$(e, \mu, \tau)$, $\langle H\rangle \simeq 174~{\rm GeV}$
is the vacuum expectation value,
$\xi(x^{}_{kj}) = \sqrt{x^{}_{kj}} \left\{1 + 1/\left(1 - x^{}_{kj}\right)
+ \left(1 + x^{}_{kj}\right) \ln \left[x^{}_{kj} / \left(1 + x^{}_{kj}\right)
\right] \right\}$ and $\zeta(x^{}_{kj}) = 1/\left(1 - x^{}_{kj}\right)$
with $x^{}_{kj} \equiv {M}^2_k/{M}^2_j$ are the loop functions.
Substituting Eq.~(\ref{7}) into Eq.~(\ref{21}), we arrive at the explicit
results for the $N^{}_4$ decays:
\newpage
\begin{align}
\varepsilon^{}_{4 e} = & \hspace{0.13cm}
\frac{1}{8\pi \langle H\rangle^2 I^{}_{44}} \Bigg\{
M^2_5 s^{}_{14} s^{}_{15} \Bigg[ \xi(x^{}_{54}) \sum^3_{i=1}
s^{}_{i4} s^{}_{i5} \sin\left(\alpha^{}_1 + \alpha^{}_i\right)
+ \zeta(x^{}_{54}) \Big[ s^{}_{24} s^{}_{25} \sin\left(\alpha^{}_1 - \alpha^{}_2\right)
\nonumber \\
&
+ s^{}_{34} s^{}_{35} \sin\left(\alpha^{}_1 - \alpha^{}_3\right)\Big] \Bigg]
- M^2_6 s^{}_{14} s^{}_{16} \Bigg[ \xi(x^{}_{64}) \sum^3_{i=1}
s^{}_{i4} s^{}_{i6} \sin\left(\gamma^{}_1 + \gamma^{}_i\right)
\nonumber \\
&
+ \zeta(x^{}_{64}) \Big[ s^{}_{24} s^{}_{26} \sin\left(\gamma^{}_1 - \gamma^{}_2\right)
+ s^{}_{34} s^{}_{36} \sin\left(\gamma^{}_1 - \gamma^{}_3\right)\Big] \Bigg] \Bigg\} \; ,
\nonumber \\
\varepsilon^{}_{4 \mu} = & \hspace{0.13cm}
\frac{1}{8\pi \langle H\rangle^2 I^{}_{44}} \Bigg\{
M^2_5 s^{}_{24} s^{}_{25} \Bigg[ \xi(x^{}_{54}) \sum^3_{i=1}
s^{}_{i4} s^{}_{i5} \sin\left(\alpha^{}_2 + \alpha^{}_i\right)
+ \zeta(x^{}_{54}) \Big[ s^{}_{14} s^{}_{15} \sin\left(\alpha^{}_2 - \alpha^{}_1\right)
\nonumber \\
&
+ s^{}_{34} s^{}_{35} \sin\left(\alpha^{}_2 - \alpha^{}_3\right)\Big] \Bigg]
- M^2_6 s^{}_{24} s^{}_{26} \Bigg[ \xi(x^{}_{64}) \sum^3_{i=1}
s^{}_{i4} s^{}_{i6} \sin\left(\gamma^{}_2 + \gamma^{}_i\right)
\nonumber \\
&
+ \zeta(x^{}_{64}) \Big[ s^{}_{14} s^{}_{16} \sin\left(\gamma^{}_2 - \gamma^{}_1\right)
+ s^{}_{34} s^{}_{36} \sin\left(\gamma^{}_2 - \gamma^{}_3\right)\Big] \Bigg] \Bigg\} \; ,
\nonumber \\
\varepsilon^{}_{4 \tau} = & \hspace{0.13cm}
\frac{1}{8\pi \langle H\rangle^2 I^{}_{44}} \Bigg\{
M^2_5 s^{}_{34} s^{}_{35} \Bigg[ \xi(x^{}_{54}) \sum^3_{i=1}
s^{}_{i4} s^{}_{i5} \sin\left(\alpha^{}_3 + \alpha^{}_i\right)
+ \zeta(x^{}_{54}) \Big[ s^{}_{14} s^{}_{15} \sin\left(\alpha^{}_3 - \alpha^{}_1\right)
\nonumber \\
&
+ s^{}_{24} s^{}_{25} \sin\left(\alpha^{}_3 - \alpha^{}_2\right)\Big] \Bigg]
- M^2_6 s^{}_{34} s^{}_{36} \Bigg[ \xi(x^{}_{64}) \sum^3_{i=1}
s^{}_{i4} s^{}_{i6} \sin\left(\gamma^{}_3 + \gamma^{}_i\right)
\nonumber \\
&
+ \zeta(x^{}_{64}) \Big[ s^{}_{14} s^{}_{16} \sin\left(\gamma^{}_3 - \gamma^{}_1\right)
+ s^{}_{24} s^{}_{26} \sin\left(\gamma^{}_3 - \gamma^{}_2\right)\Big] \Bigg] \Bigg\} \; .
\label{22}
\end{align}
A sum of these three CP-violating asymmetries, denoted as $\varepsilon^{}_{4} \equiv
\varepsilon^{}_{4 e} + \varepsilon^{}_{4 \mu} + \varepsilon^{}_{4 \tau}$, is usually
referred to as the flavor-independent asymmetry. The expression of
$\varepsilon^{}_{4}$ turns out to be independent of the loop functions $\zeta(x^{}_{54})$
and $\zeta(x^{}_{64})$:
\begin{align}
\varepsilon^{}_{4} = & \hspace{0.13cm}
\frac{1}{8\pi \langle H\rangle^2 I^{}_{44}} \Bigg[
\sum^3_{i=1} \sum^3_{i^\prime = 1} s^{}_{i4} s^{}_{i^\prime 4}
\Big[M^2_5 \xi(x^{}_{54}) s^{}_{i5} s^{}_{i^\prime 5}
\sin\left(\alpha^{}_i + \alpha^{}_{i^\prime}\right)
- M^2_6 \xi(x^{}_{64}) s^{}_{i6} s^{}_{i^\prime 6}
\sin\left(\gamma^{}_i + \gamma^{}_{i^\prime}\right)\Big]\Bigg] \; .
\label{23}
\end{align}
A salient feature of the results in Eqs.~(\ref{22}) and (\ref{23}) is that the
phase parameters $\beta^{}_i$ (for $i = 1, 2, 3$) do not contribute to
CP violation in the $N^{}_4$ decay modes.

It is straightforward to work out the CP-violating asymmetries
$\varepsilon^{}_{5 \alpha}$ and $\varepsilon^{}_{6 \alpha}$ (for $\alpha = e, \mu, \tau$)
in the same way. We find that the analytical expressions of $\varepsilon^{}_{5 \alpha}$
and $\varepsilon^{}_{5}$ can be easily written out from those of
$\varepsilon^{}_{4 \alpha}$ and $\varepsilon^{}_{4}$
with the subscript replacements $\left(4, 5, 6\right) \to
\left(5, 4, 6\right)$ together with $\left(\alpha^{}_i , \gamma^{}_i\right)
\to -\left(\alpha^{}_i , \beta^{}_i \right)$; and similarly the expressions of
$\varepsilon^{}_{6 \alpha}$ and $\varepsilon^{}_{6}$ can be directly read off from
those of $\varepsilon^{}_{4 \alpha}$ and $\varepsilon^{}_{4}$ with the help of the
subscript replacements $\left(4, 5, 6\right) \to \left(6, 5, 4\right)$ together with
$\left(\alpha^{}_i , \gamma^{}_i\right) \to -\left(\beta^{}_i , \gamma^{}_i \right)$.

As a result, CP violation in the decays of three heavy Majorana neutrinos only
involve the following 27 linear combinations of the 6 original seesaw phase parameters:
\begin{equation}
\begin{array}{lllllllll}
\sin 2\alpha^{}_1 , & \sin 2\alpha^{}_2 , & \sin 2\alpha^{}_3 ;
& \sin 2\beta^{}_1 , & \sin 2\beta^{}_2 , & \sin 2\beta^{}_3 ;
& \sin 2\gamma^{}_1 , & \sin 2\gamma^{}_2 , & \sin 2\gamma^{}_3 ;
\nonumber
\end{array}
\end{equation}
\begin{equation}
\begin{array}{llllll}
\sin\left(\alpha^{}_1 \pm \alpha^{}_2\right) ,
& \sin\left(\alpha^{}_2 \pm \alpha^{}_3\right) ,
& \sin\left(\alpha^{}_3 \pm \alpha^{}_1\right) ;
& \sin\left(\beta^{}_1 \pm \beta^{}_2\right) ,
& \sin\left(\beta^{}_2 \pm \beta^{}_3\right) ,
& \sin\left(\beta^{}_3 \pm \beta^{}_1\right) ;
\nonumber \\
\sin\left(\gamma^{}_1 \pm \gamma^{}_2\right) ,
& \sin\left(\gamma^{}_2 \pm \gamma^{}_3\right) ,
& \sin\left(\gamma^{}_3 \pm \gamma^{}_1\right) .
&&&
\nonumber
\end{array}
\end{equation}
One can see that each of the three $\varepsilon^{}_j$ asymmetries (for $j = 4, 5, 6$)
depends only on 12 simple phase combinations, and each of the nine
$\varepsilon^{}_{j\alpha}$ asymmetries (for $j = 4, 5, 6$ and $\alpha = e, \mu, \tau$)
is dependent only upon 10 simple phase combinations. It is certainly allowed
to compactly express $\varepsilon^{}_{j\alpha}$ and $\varepsilon^{}_j$ as
\begin{align}
\varepsilon^{}_{j\alpha} = & \hspace{0.1cm}
\sum^3_{i=1} \left(C^{\prime}_{\alpha i} \sin\alpha^{}_i +
C^{\prime}_{\beta i} \sin\beta^{}_i\right) \; ,
\nonumber \\
\varepsilon^{}_{j~} = & \hspace{0.1cm}
\sum^3_{i=1} \left(C^{\prime\prime}_{\alpha i} \sin\alpha^{}_i +
C^{\prime\prime}_{\beta i} \sin\beta^{}_i\right) \; ,
\label{24}
\end{align}
where the coefficients $C^{\prime}_{\alpha i}$, $C^{\prime}_{\beta i}$,
$C^{\prime\prime}_{\alpha i}$ and $C^{\prime\prime}_{\beta i}$ (for $i = 1, 2, 3$)
can be easily extracted from the expressions of $\varepsilon^{}_{j\alpha}$ and
$\varepsilon^{}_j$ (for $j = 4, 5, 6$ and $\alpha = e, \mu, \tau$) obtained above.
A comparison between Eqs.~(\ref{19}) and (\ref{24}) indicates that a direct
relationship between ${\cal J}^{}_\nu$ and $\varepsilon^{}_{j\alpha}$ (or
$\varepsilon^{}_j$) can {\it in principle} be established.

\section{Conclusion}

It is true that ``for neutrino masses, the considerations have always been qualitative,
and despite some interesting attempts, there has never been a convincing quantitative
model of the neutrino masses", as remarked by Edward Witten in 2001~\cite{Witten:2000dt}.
One of the main reasons for this unfortunate situation is that none of the theoretically
well-motivated mechanisms, including the elegant canonical seesaw mechanism under
discussion, is capable of determining its flavor structure from the very beginning.
One possible way out is certainly to invoke possible lepton flavor symmetries~\cite{Xing:2020ijf,Xing:2015fdg,King:2017guk,Feruglio:2019ybq,Xing:2022uax,
Ding:2023htn,Ding:2024ozt}, but so far none of them stands out as a convincing basis of
model building.

In this case we find that a full Euler-like block parametrization of the flavor structure
in the seesaw mechanism~\cite{Xing:2007zj,Xing:2011ur} can help a lot because it
makes some important issues {\it analytically calculable}. The present work gives a good
example of this kind --- a general and explicit connection between CP violation in the
flavor oscillations of three light neutrino species and that in the lepton-number-violating
decays of three heavy Majorana neutrinos is {\it for the first time} established via the
seesaw bridge. Given that the canonical seesaw mechanism is the most
natural, economical and convincing theory for neutrino mass generation and baryogenesis via
leptogenesis, we expect that our model-independent result will pave the way for probing
the cosmic matter-antimatter asymmetry from CP violation in neutrino oscillations at low
energies.

We conclude that there is surely a {\it direct} connection between the high- and low-scale
effects of CP violation in the seesaw framework, although this seesaw-bridged connection
is rather complicated and thus sets a big challenge to its experimental testability.

Encouraged by this {\it proof-of-concept} work, we are going to calculate all
the {\it derivational} flavor parameters --- three masses, three flavor
mixing angles and three CP-violating phases of the light Majorana neutrinos
by using the {\it original} flavor parameters in the canonical seesaw
mechanism~\cite{Xing-Zhu2024}, such that its parameter space can be constrained
by both more precise neutrino oscillation data and more accurate
non-oscillation measurements in the near future. Instead of testing many specific
seesaw models which are based on various special assumptions, our approach will
offer a model-independent way to test the canonical seesaw mechanism as a whole
although the number of observable quantities at low energies cannot match the
number of the original seesaw flavor parameters. Such an attempt is fundamentally
meaningful in the era of precision measurements.

\section*{Acknowledgements}

I am greatly indebted to Jun Cao, Yifang Wang, Di Zhang, Shun Zhou and Jing-yu Zhu
for asking the thought-provoking question and for many enlightening discussions,
to Shun Zhou for sharing his deep insight on CP violation with me, and especially
to Jing-yu Zhu for her kindly checking my {\it handwork} calculations with
{\it Mathematica}, correcting my typing mistakes associated with a part of the
expression of $T^{}_{222}$ and reading through the revised version of this paper. 
I would also like to thank Wilfried Buchm$\rm\ddot{u}$ller for
some pleasant discussions about leptogenesis on the mountains of Oberw$\rm\ddot{o}$lz
in September 2023. This work was supported in part by the National Natural Science
Foundation of China under grant No. 12075254.


\end{document}